\documentclass[]{tPHM2e}

\usepackage{hhline}
\def\urhh{U(Ru$_{0.96}$Rh$_{0.04}$)$_2$Si$_2$}
\def\urfs{U(Ru$_{1-x}$Fe$_{x}$)$_2$Si$_2$}
\def\urs{URu$_2$Si$_2$}

\begin{document}
\doi{10.1080/14786435.20xx.xxxxxx}
\issn{1478-6443}
\issnp{1478-6435}
\jvol{00} \jnum{00} \jyear{2010} 

\markboth{Frederic Bourdarot, Stephane Raymond and Louis-Pierre Regnault}{Philosophical Magazine}

\articletype{PREPRINT}

\title{Neutron~scattering~studies~on~\urs.}

\author{Frederic Bourdarot$^{\rm a}$$^{\rm b}$$^{\ast}$\thanks{$^\ast$Corresponding author. Email: frederic.bourdarot@cea.fr
\vspace{6pt}}, Stephane Raymond$^{\rm a}$$^{\rm b}$ and Louis-Pierre Regnault$^{\rm a}$$^{\rm b}$\\\vspace{6pt}  $^{\rm a}${\em{Univ. Grenoble Alpes, INAC-SPSMS, F-38000 Grenoble, France}};\\ $^{\rm b}${\em{CEA, INAC-SPSMS, F-38000 Grenoble, France}}\\\vspace{6pt}\received{v1.0 released \today}}

\maketitle

\begin{abstract}
This paper is aiming to review some of the neutron scattering studies performed on \urs\ in Grenoble. This compound has been studied for a quarter of century because of a so-called hidden order ground state visible by most of the bulk experiments but almost invisible by microscopic probes like neutrons, muons NMR or x-ray. We stress on some aspects that were not addressed previously.

Firstly, the comparison of the cell parameters in the 1-2-2 systems seems to point that the magnetic properties of \urs\ are leading by an U$^{4+}$ electronic state. Secondly, a compilation of the different studies of the tiny antiferromagnetic moment indicates that the tiny antiferromagnetic moment has a constant value which may indicate that it is not necessary extrinsic. We also present the last development on the magnetic form factor measurement in which the magnetic density rotates when entering in the hidden order state. To end, the thermal dependence of the two most intense magnetic excitation at \textbf{Q$_0$}=(1,0,0) and \textbf{Q$_1$}=(0.6,0,0) seems to indicate two different origins or processes for these excitations.
.\bigskip

\bigskip

\end{abstract}

\section{Introduction.}

The uranium and uranium-based materials are an endless source of unconventional and exotic physical properties: their \textit{5f} electrons are intermediate between itinerant and localized and they are exposed to several interactions (exchange, spin-orbit and Coulomb, ...) without clean-cutting hierarchy. Associated with the large orbital angular moments of these electrons, there is also the possibility for high order multipolar electromagnetic asphericities to form and to order~\cite{Santini:2009}.

The tetragonal heavy fermion superconductor compound URu$_2$Si$_2$ is a good example of such materials and has puzzled physicists for more than two decades~\cite{Mydosh:2011a}. Due to the dual character of $5f$ electrons in \urs\, between localized (leading to the possibility of multipolar ordering) and itinerant (possibility of large Fermi surface instabilities), this compound is a key example which has been the subject of a large variety of experiments\cite{amitsuka:2007}.
\begin{figure}[!h]
\begin{center}
\includegraphics[width=100mm]{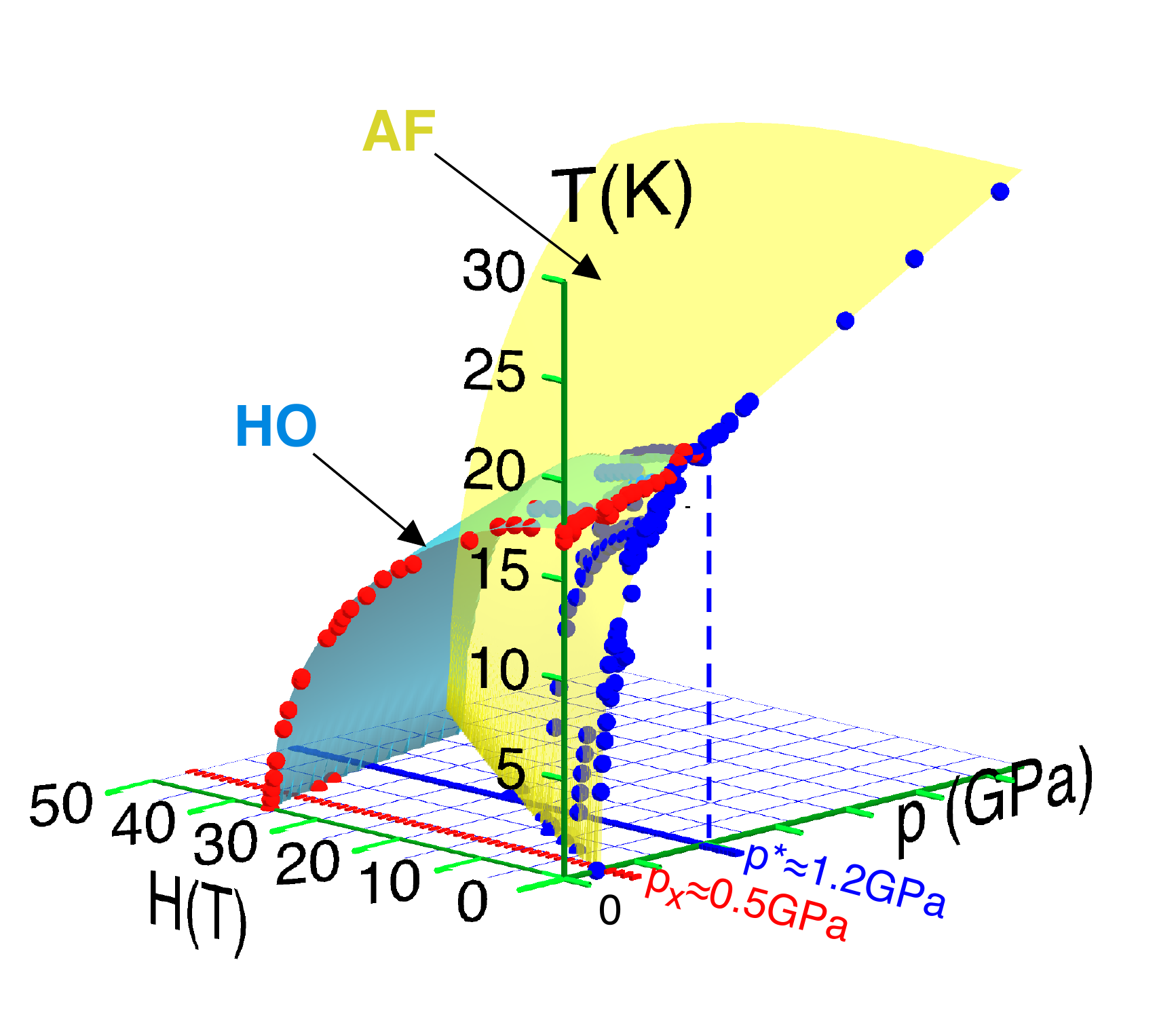}
\caption{Schematic phase diagram $(T,H,P)$ of \urs. The data come from Ref.\cite{Hassinger:2008,Aoki:2009,Aoki:2010,Jaime:2002}. P$_X\simeq0.5$ GPa corresponds to the critical pressure, P$^*\simeq1.4$ GPa to the pressure where \urs\ switches directly from PM state to AF state. The superconducting phase is not presented to simplify the phase diagram.}
\label{fig1}
\end{center}
\end{figure}
The mysterious phase transition at T$_0 \sim$ 17.8 K of this $5f$ heavy-electron compound is characterized by large bulk anomalies and sharp magnetic excitations in q-space and energy, at different \textbf{Q}-vectors (\textbf{Q$_0$}=(1,0,0) and \textbf{Q$_1$}=(0.6,0,0)). Concomitant with this order, a tiny but persistent antiferromagnetic moment ($\sim$ 0.02 $\mu_B$) is measured in all samples by elastic neutron scattering~\cite{Broholm:1987,Broholm:1991} and x-ray scattering~\cite{Isaacs:1990} with a wave-vector \textbf{Q$_{AF}$}=(0,0,1) (equivalent to \textbf{Q$_0$}). It is difficult to consider this tiny staggered moment as the order parameter in a conventional antiferromagnetism frame as it cannot be reconciled with the jump which is observed in the specific heat $\Delta C/T \sim$ 300 mJ/K$^2$mol, involving an entropy S $\sim$ 0.2R$\ln$(2). Because there is no determination of the order parameter (OP), the order in \urs\ is named the hidden order (HO). However, under pressure, \urs\ orders in a high-moment antiferromagnetic (AF) structure with the wave-vector \textbf{Q$_{AF}$} and a moment of 0.36-0.4~$\mu_B$\cite{Bourdarot:2004b,Bourdarot:2005}. The well-defined phase diagram (see Fig.\ref{fig1}) shows that when \urs\ switches from HO to AF state at a critical pressure P$_X\simeq$~0.5~GPa, the bulk superconductivity disappears \cite{Hassinger:2008,Hassinger:2008a} as well as the antiferromagnetic excitation E$_0$ at \textbf{Q$_0$}, signature of the HO phase \cite{Villaume:2008}. At P$_X$, the excitation E$_1$ at \textbf{Q$_1$} jumps from 5~meV to 8~meV\cite{Villaume:2008}. The HO-AF boundary $T_X(P)$ seems to meet the $T_0(P)$ line at the tricritical point ($T^\star\simeq$19.3~K, $P^\star\simeq$1.36~GPa) \cite{Hassinger:2008}; above $P^\star$, a unique ordered phase (AF) is achieved under pressure below $T_N(P)$. Previous NMR experiments \cite{matsuda:2001,kohara:1986} as well as transport measurements \cite{maple:1986,Hassinger:2008} indicate clearly that nesting occurs at $T_0(P)$ as well as at $T_N(P)$. The occurrence of nesting which triggers a sudden drop of the density of states at the Fermi level, certainly plays a key role in the recovery of localized properties for both HO and AF phases. Under magnetic field (applied along \textbf{c}-axis), the pressurized AF phase is unstable and \urs\ reenters into the HO state\cite{Aoki:2009,Aoki:2010}(see Fig.\ref{fig1}). 
The order parameter is not yet determined: spin or charge density wave~\cite{maple:1986, mineev:2005}, multipolar ordering~\cite{santini:1994, ohkawa:1999, hanzawa:2007, kiss:2005,Cricchio:2009,Ikeda:2012}, orbital antiferromagnetism~\cite{chandra:2002}, chiral spin state \cite{gorkov:1992}, nematic order~\cite{Fujimoto:2011}, helicity order~\cite{varma:2006}, and hastatic order~\cite{Chandra:2013} have been proposed.
Recent attempts to search for a possible quadrupolar order by resonant x-ray scattering have failed~\cite{Amitsuka:2010,Walker:2011}.
Thermal expansion measurements established that stress will increase T$_0$ when it is applied along the \mbox{\textbf{a}-axis} and decrease T$_0$ when it is applied along the \mbox{\textbf{c}-axis}. The opposite effect is observed for the evolution of superconductivity\cite{Bakker:1992,Guillaume:1999}. Recently a theoretical model \cite{Harima:2010} indicated that the space groups of the paramagnetic (PM) state and of the HO state are different but may keep unchanged the atomic positions, this explaining why most of local probes did not detect any modification in the crystal structure.

\section{Electronic state of UR\lowercase{u}$_2$S\lowercase{i}$_2$.}
\label{PR1}

\begin{figure}[h]
\begin{center}
\includegraphics[width=100mm]{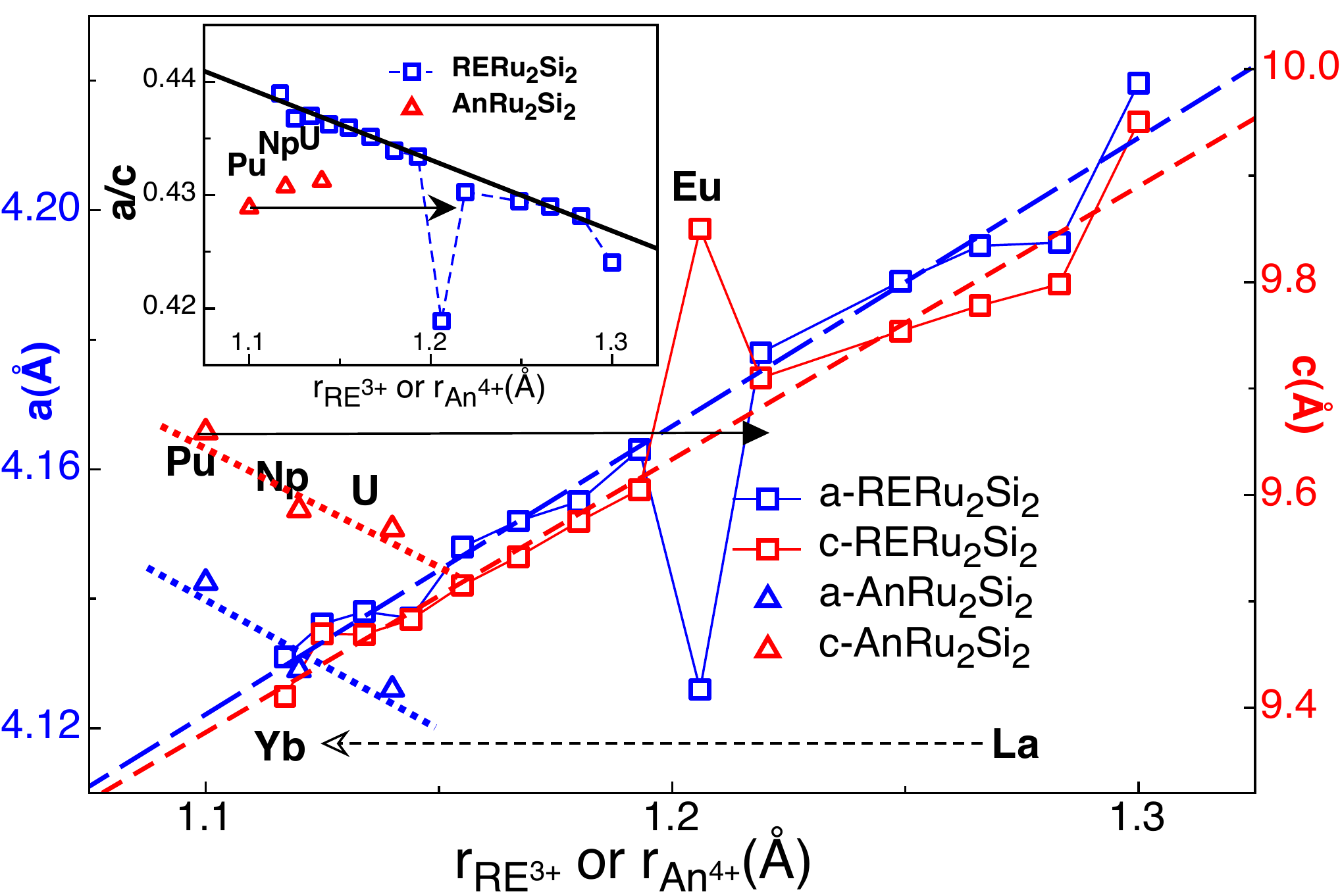}
\caption{Variation of the tetragonal lattice parameters $a$ and $c$ (and of the ratio $a/c$ in the inset) of RERu$_2$Si$_2$ compounds (with RE = Rare-Earth) or AnRu$_2$Si$_2$ (with An = Actinide) versus of the ionic radius.}
\label{fig2}
\end{center}
\end{figure}

As in many heavy fermion compounds, the electronic state of \urs\ is not clearly defined and the calculation of the electronic bands does not give a clear answer as well. However, their results seem to indicate that there are two contributions near the Fermi surface, one corresponding to a localized contribution with two electrons and a second with a quite broad contribution with 0.7 electrons\cite{Ikeda:2012}. There is then a clear competition between localized and itinerant magnetism in \urs\ and the variation of the lattice parameters $a$ and $c$ of the compounds XRu$_2$Si$_2$ (where X can be a Rare-Earth or an Actinide elements) may enlighten this competition. For Rare-Earth compounds, the lattice parameters $a$ and $c$ increase linearly with the radius size of the Rare-Earth proving that the Rare-Earth keeps for all the series the valency 3$^+$ (excepted Eu as shown in figure~\ref{fig2}). For Actinide compounds, the contraction of the lattice parameters when the isoelectronic ion size increases as the ions go from plutonium to uranium, proves a large modification of the electronic state or an increase of the hybridization: while plutonium valency in PuRu$_2$Si$_2$ can be considered as almost 3$^+$, \urs\ has, according to the lattice variation almost a 4$^+$ valency or a large hybridization of about one $5f$ electronic band in the U$^{3+}$ state. This result is the starting point of the treatment of the U ion in the configuration 5$f^2$ for \urs.

The (U$^{4+}$ or $5f^2$) electronic configuration of uranium ion is in agreement with measurements of the magnetic excitations at \textbf{Q$_0$} and \textbf{Q$_1$} under magnetic field. Indeed these excitations show no-splitting under a magnetic field applied along the $c$-axis~\cite{Santini:2000,Bourdarot:2003} or in the (a,b) plane~\cite{Bourdarot:2013}. This result proves that the uranium ion can not be a Kramers ion  ($5f^3$) nor a non-Kramer ion with a doublet as ground state: doublet being the largest degeneracy for U$^{4+}$ in a tetragonal symmetry (see appendix~\ref{Ann2}).

\section{Neutron diffraction.}

\begin{figure}[!h]
\begin{minipage}{.475\linewidth}
\begin{center}
\includegraphics[width=66mm]{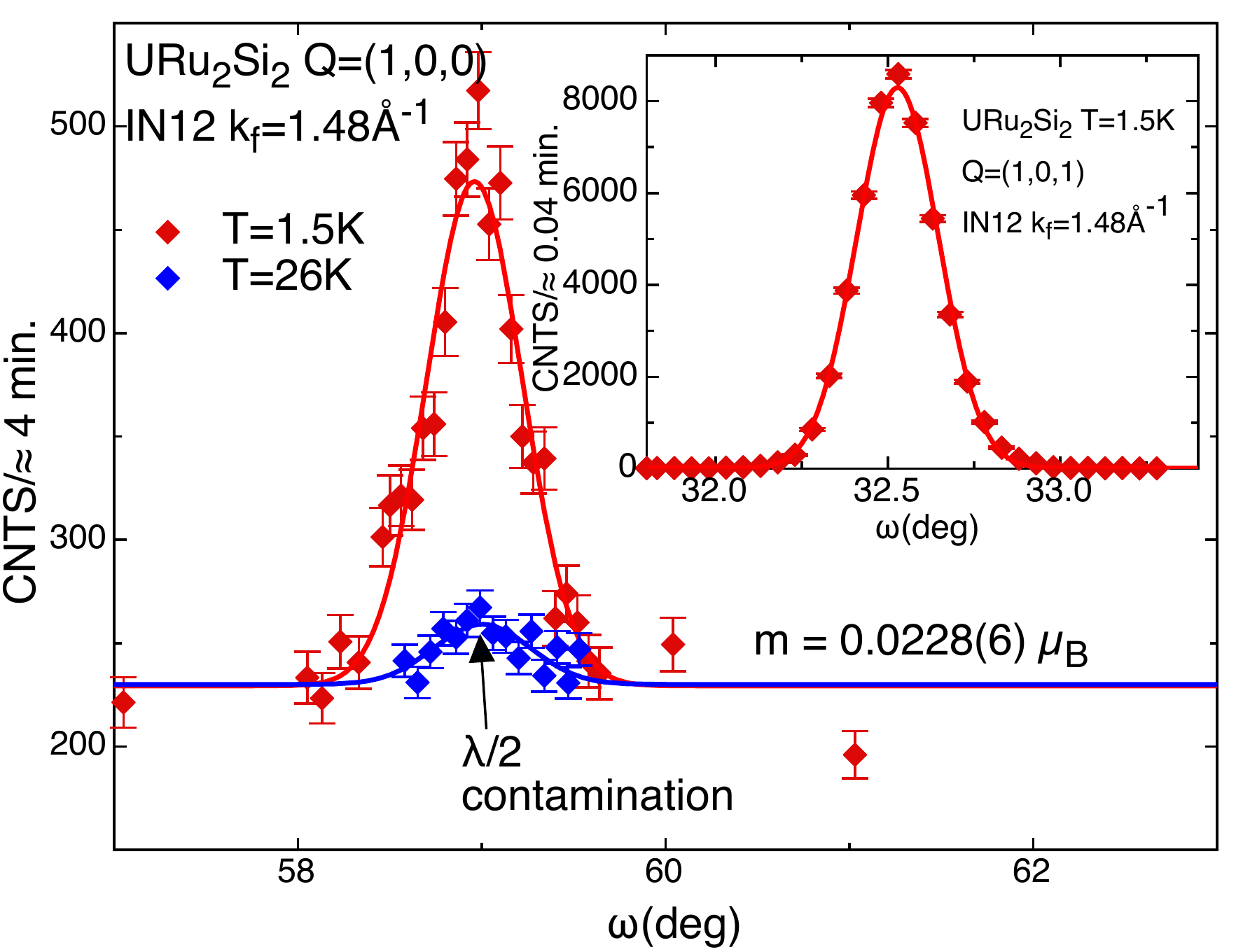}
\caption{$\omega$-scan of the magnetic peak at Q$_0$=(1,0,0) on IN12. In inset comparison with the weak nuclear peak (1,0,1).}
\label{L(1,0,0)}
\end{center}
\end{minipage}
\hfill
\begin{minipage}{.475\linewidth}
\begin{center}
\includegraphics[width=66mm]{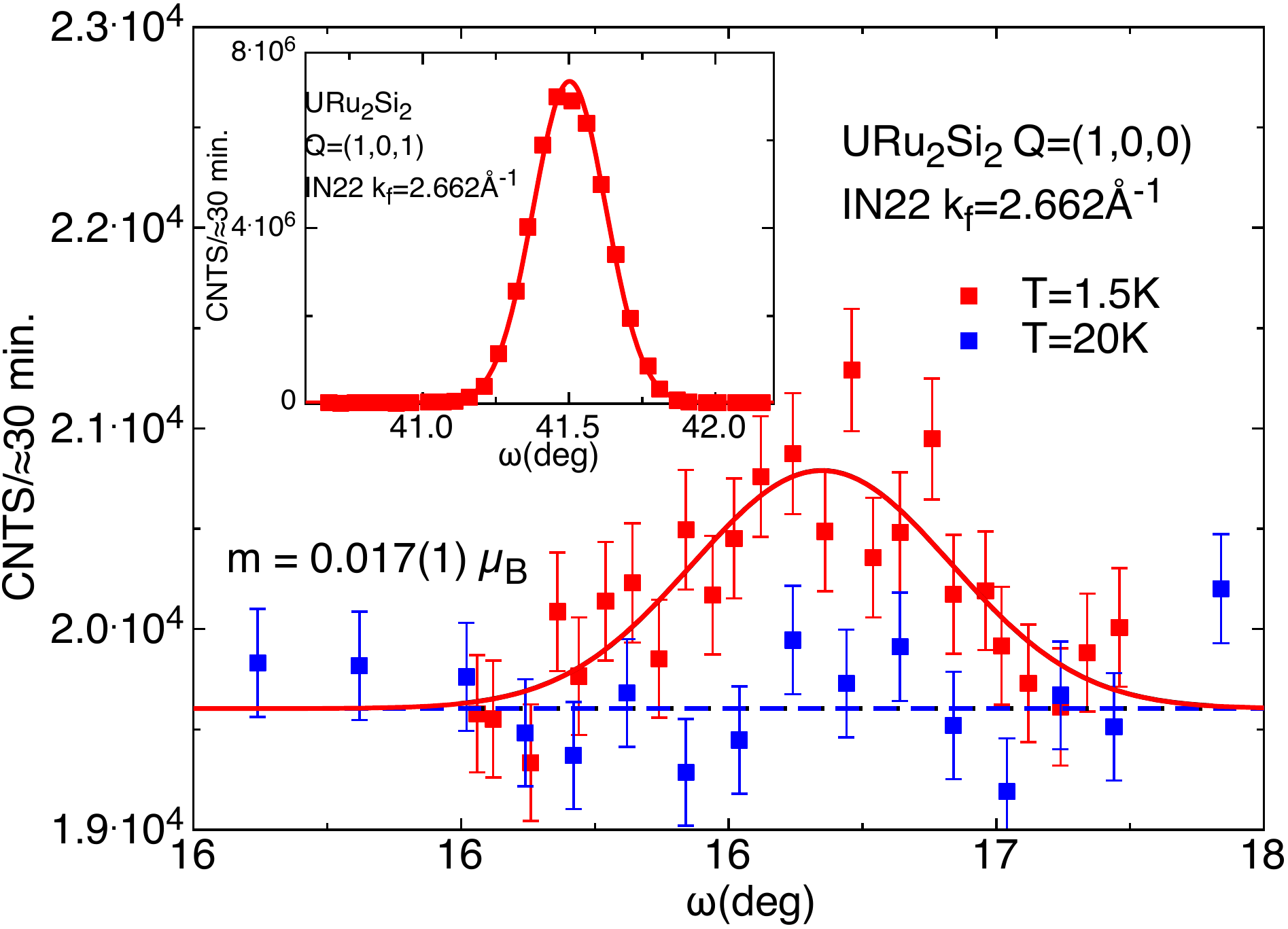}
\caption{$\omega$-scan of the magnetic peak at Q$_0$=(1,0,0) on IN22.  In inset comparison with the weak nuclear peak (1,0,1).}
\label{L(1,0,0)2}
\end{center}
\end{minipage}
\end{figure}

The study of the antiferromagnetic moment at \textbf{Q$_0$} was performed many times, from C.~Broholm~\cite{Broholm:1987,Broholm:1991} in the 1990s to P.G.~Niklowitz and N.~Metoki~\cite{Niklowitz:2010,Metoki:2013} very recently, with values ranging from 0.012 to 0.03~$\mu_B$. We have also determined this tiny moment value on a large crystal on the cold neutron three-axis spectrometer (TAS) IN12 ($k_f=1.5$\AA$^{-1}$), and by using the thermal neutron three-axis spectrometer IN22, ($k_f=2.662$\AA$^{-1}$) with a small crystal on which we had previously determined the nuclear structure and the extinction (on the two-axis lifting arm detector D23 from the study of the magnetization density map\cite{Ressouche:2012}). The value of the moment found on IN12 is 0.0228(6)~$\mu_B$, while it is only 0.017(1)~$\mu_B$ on IN22. These values are determined from data without extinction corrections  (figures~\ref{L(1,0,0)} and \ref{L(1,0,0)2}, respectively) and are overestimated due to the very high quality of the crystal leading to a large extinction.
\begin{figure}[!h]
\begin{center}
\includegraphics[width=100mm]{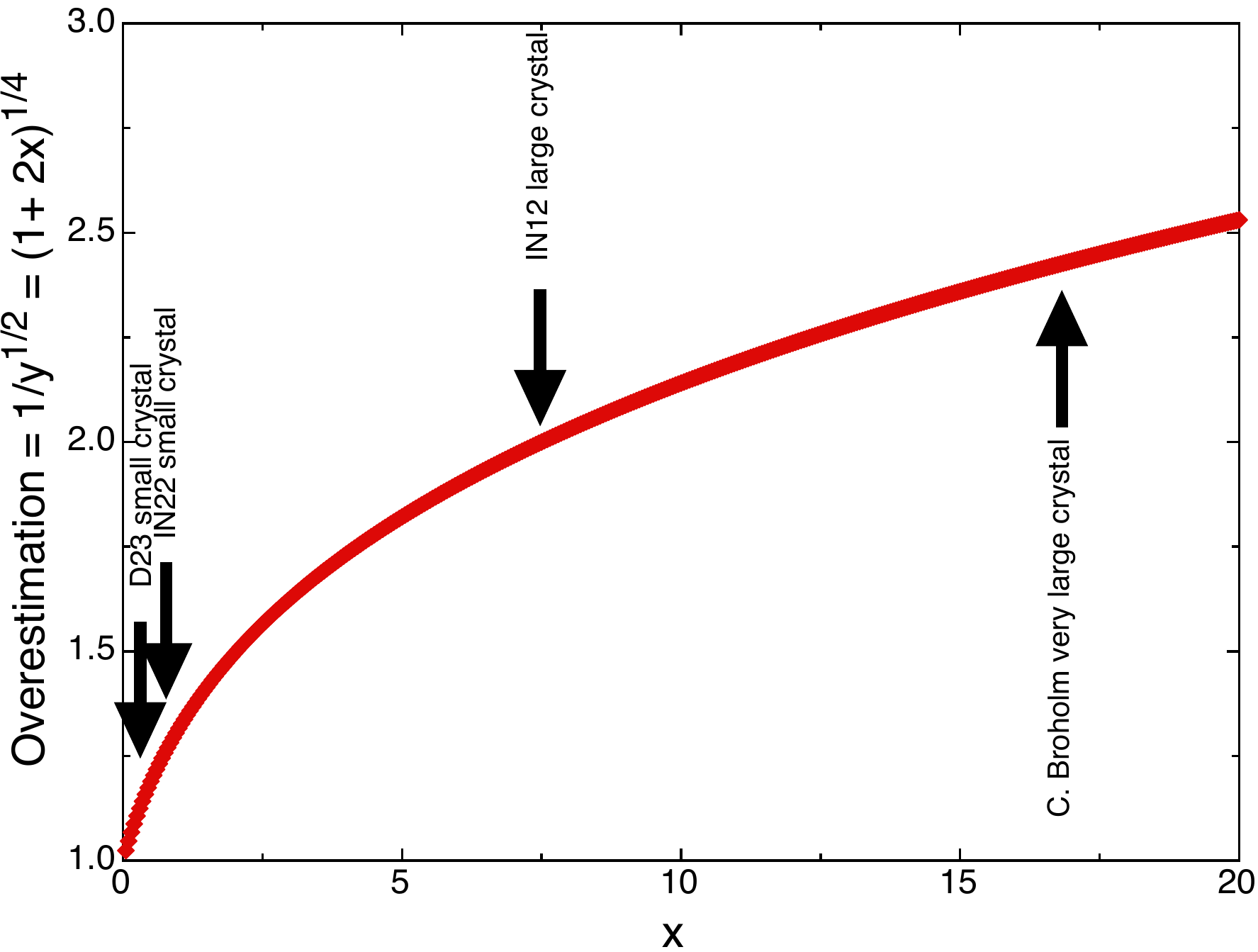}
\caption{Overestimation of the antiferromagnetic moment due to extinction of a weak nuclear reflection as (1,0,1) or (1,1,0) as a function of $x=\frac{\lambda^3|F_N|^2 T}{2\sqrt{\pi}V_0^2\sin(2\theta)\eta}$, $\eta$ is the crystal mosaicity and $T$ the path of the beam in the crystal.}
\label{yx}
\end{center}
\end{figure}
To determine the value of this small antiferromagnetic moment in the case of \urs, we have used the relation:
\begin{equation}
\label{IL}
m_0\cdot f\cdot p = \frac{F_N(h,k,l)}{\sqrt{\langle \sin^2 \alpha \rangle}} \sqrt{\frac{I/L_{mag}}{I/L_{nuc}(h,k,l)}}
\end{equation}
where $p$=0.2696 $\cdot 10^{-12}$ cm, $f$ is the magnetic form factor, $\alpha$ the angle between the scattering vector and the direction of the magnetic moment ($c$ in the case of \urs) and $L$ the Lorentz factor. This relation gives good results when $I/L_{mag} \sim I/L_{nuc}(h,k,l)$ or when extinction is low for both intensities. For \urs, as shown in figures~\ref{L(1,0,0)} and \ref{L(1,0,0)2}, there are three orders of magnitude between the less intense nuclear peaks and the magnetic ones. The extinction of the nuclear Bragg peaks can be adjusted by the expression $I/L_{nuc} \propto y(x) F^2_N(h,k,l)$ where $y(x)=1/\sqrt{1+2x}$ with $x=\frac{\lambda^3|F_N|^2 T}{2\sqrt{\pi}V_0^2\sin(2\theta)\eta}$, $\eta$  is the crystal mosaicity and $T$ the beam path inside the crystal ( the expression is ordinarily valid for values of $x\leqslant5$). The extinction was precisely determined on a small sample using the diffractometer D23 with a wavelength $\lambda$=1.227\AA. For the weak nuclear Bragg peak (1,0,1) we already obtain an extinction $y(\lambda$=1.227\AA)$\sim$0.85. Using the same parameters, the extinction in the case of large crystals and longer wavelengths, and the overestimation of tiny ordered moment  m$_0$ were determined. This overestimation as a function of $x$ is giving by $1/\sqrt{y}=(1+2x)^{1/4}$ and it is represented for different experiments on the figure~\ref{yx}.
Taking into account this overestimation, we deduce that the antiferromagnetic moment value of m$_0$ is almost constant (0.012~$\mu_B$) for the different origin of the samples.
The extinction corrections shows as well that estimations of the antiferromagnetic moment using for the normalization of intensities strong nuclear reflections such as the (2,0,0) or (0,0,4) which have $x$ of hundreds, does not make sense.


Assuming the tiny antiferromagnetic moment of 0.012(1)~$\mu_B$ extrinsic and originating from point defects of the high-pressure antiferromagnetic phase, a volume of defects corresponding to 2.5/1000 of the total volume or 1 magnetic atom in a volume of 32500~ \AA$^3$ can be deduced. These values seem low to produce correlated magnetic domains.

In conclusion, for all the measurements performed by elastic neutron scattering, we have shown that the tiny antiferromagnetic ordered moment at \textbf{Q$_0$} is not only always present for any samples but also its value of m$_0$ 0.012$\mu_B$ appears to be sample independent after appropriate extinction corrections.

The large extinction is confirmed by the Neutron Larmor diffraction results that give a distribution of lattice parameters almost identical to the distribution found in a perfect silicon single crystal\cite{Bourdarot:2011}. However, even if the persistence of the tiny antiferromagnetic moment value seems to indicate an intrinsic feature, the neutron results cannot prove it. Many other experiments should have seen this tiny ordered moment as well, which is not the case. For example, in a NMR experiment with a dipolar moment of 0.012~$\mu_B$ on the uranium site, it should appear a magnetic field of $\simeq$17G at the silicon site. However, this magnetic field may be affected by the magnetic contribution of the HO.


\section{Magnetic form factor.}

\begin{figure}[!h]
\begin{minipage}{.48\linewidth}
\begin{center}
\includegraphics[width=50mm]{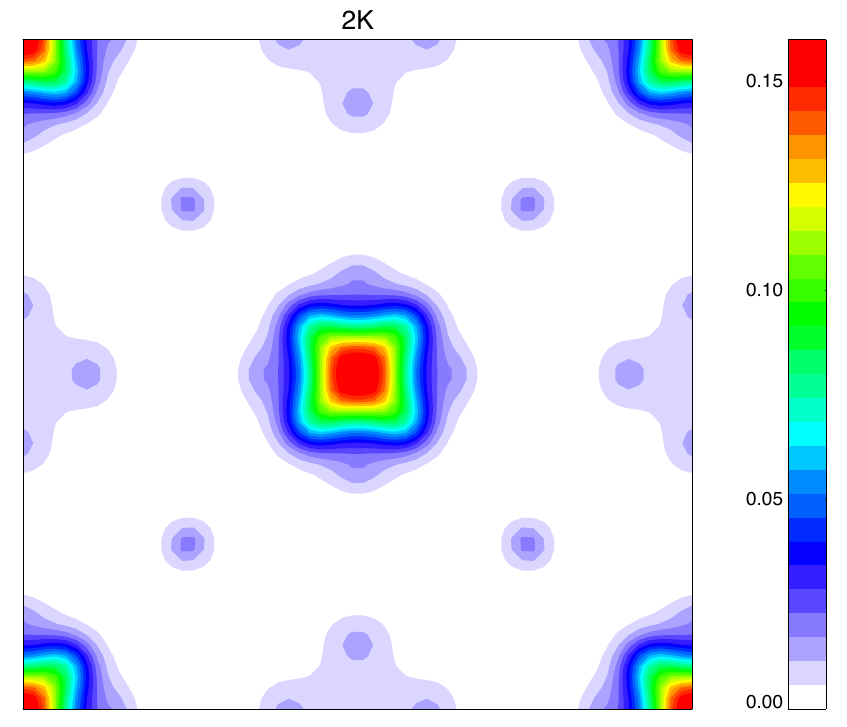}
\end{center}
\end{minipage}
\hfill
\begin{minipage}{.48\linewidth}
\begin{center}
\includegraphics[width=50mm]{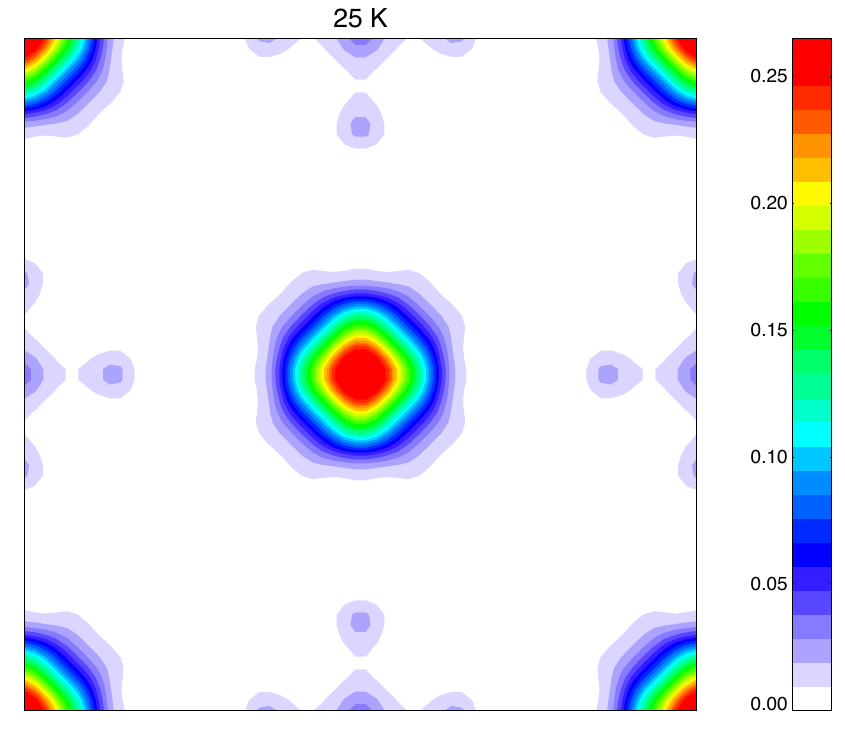}
\end{center}
\end{minipage}
\caption{Projection of the MaxEnt reconstructed magnetization distribution in \urs. Left along the [001] axis at T = 2 K (below T$_0$). Right: same MaxEnt projection at T = 25 K (above T$_0$). The scale is giving in $\mu_B$/(unit cell).}
\label{Rf1}
\end{figure}

In a recent article~\cite{Ressouche:2012}, the measurement of the magnetic structure factor has revealed a modification of the "squared" field-induced magnetization distribution between the paramagnetic state and the hidden order state corresponding to a rotation of 45$^\circ$ around the vertical $c$-axis (see figure~\ref{Rf1}). The analysis of this magnetization density distribution has been performed using a localized magnetic model as described in the reference~\cite{Ressouche:2012} with a form factor developed as:
\begin{equation}
F_{\rm M}({\bf Q}) = (1+(-1)^{h+k+l}) ~B(T) ~\sin^{-2}(\alpha_{\bf Q}) ~E_{Q}
\end{equation}
where $B(T)$ is the temperature Debye-Waller factor and $\alpha_ {\bf Q}$ is the angle between the diffusion vector of the nuclear Bragg peak $ {\bf Q} = [h, k, l] $ and the magnetization axis ($c$-axis is the case of \urs). $E_Q$ is the the projection of the vectorial component of the magnetic form factor along the $c$-axis. Its definition is given in reference~\cite{Lovesey:1969}.

As described in reference~\cite{Ressouche:2012}, the crucial point is the choice of the wave-function basis. From the analyses of magnetic excitations under magnetic field, specific heat and susceptibility measurements, it has be shown that both the fundamental ground state level and the first excited level are singlets and these levels have to be looked for in the basis  $\{ \Gamma_{t1}^{(1)},\ \Gamma_{t2},\ \Gamma_{t1}^{(2)}\}$ of  U$^{4+}$ in a tetragonal symmetry. The other levels are supposed to be far away and have no influence on the magnetic properties for temperatures much lower than 150K in agreement with the specific heat analysis where  the third singlet level is estimated to be around 200K. The localized character that is shown by the extension of the magnetization distribution, similar to measurements found in rare-earth compounds. This is in agreement with the phase diagram of \urfs\ obtained by Maple's group\cite{Kanchanavatee:2012} and the phase diagrams of \urs\ under pressure that reproduce the Doniach phase diagram of cerium compounds under low pressure where localized character dominates. Contrary to this localized character, results of ARPES seems to incline towards an itinerant model~\cite{Kawasaki:2011,Boariu:2013}. Even if this duality exists indubitably in \urs, neutron diffraction is mainly sensitive to localized electrons this justifying the development of a localized model in order to explain the field-induced magnetization distribution.

The ground state wave function has been decomposed into a complex linear combination of the wave functions $\{ \Gamma_{t1}^{(1)},\ \Gamma_{t2},\ \Gamma_{t1}^{(2)}\}$ (these wave functions of the crystal electric field are developed in the annex~\ref{Ann2}):
\begin{equation}
\label{wavefunction} 
\psi = \cos\alpha\,\big\{\cos\varphi\,\Gamma_{t1}^{(1)} + \sin\,\varphi e^{i\beta_1} \Gamma_{t1}^{(2)} \big\} +\,\sin\alpha\, e^{i\beta_2} \Gamma_{t2}
\end{equation}
As this function is normalized and defined with an unknown complex phase, four parameters are needed to describe it. 

In the reference~\cite{Ressouche:2012}, it was explained that the modifications of the magnetic density maps measured above and below the transition temperature T$_0\simeq$18K can be attributed to a dotriacontapolar order $D_{4s}$. However, this proposal is ruled out as it can be easily seen that  $ \langle \psi | D_{s4} | \psi \rangle \propto \cos^2\alpha \sin \beta_1 \sin 2\varphi=0$ if $\beta_1=0$. In fact, the solution reproducing the magnetization distribution is not unique. The analysis of the experiment is much more complex than thought from the beginning and shown in the reference~\cite{Khalyavin:2014}. The analysis is then still under treatment and it has to take into account not only the symmetry of the system under magnetic field but also the results of other experiment as the specific heat, susceptibility and magnetic excitation measurements to get the answer. One possible candidate from the new analysis seems to be the dotriacontapole D$_z^\beta$, but this result needs to be confirmed.

\section{Inelastic neutron scattering}

Since the first measurements of C.~Broholm\cite{Broholm:1987,Broholm:1991}, the magnetic dispersion of \urs\ has been described as displaying two minimums, firstly at the antiferromagnetic position \textbf{Q$_0$}=(0,0,1) and secondly at the position \textbf{Q$_1$}=(0.4,0,1) which corresponds to an edge of the first Brillouin zone. However, the temperature dependence of these two excitations seems to indicate that the description as a unique excitation dispersion could be misleading, as it will be shown in the following.

\subsection{Q$_0$=(1,0,0) magnetic excitation.}

The wave-vector \textbf{Q$_0$}=(1,0,0) is equivalent to the one at \textbf{Q}=(0,0,1) in the first Brillouin zone. Because of the neutron selection rules (neutron scattering probes only the component of the magnetic moment perpendicular to the scattering vector), longitudinal magnetic signals cannot be measured at  \textbf{Q}=(0,0,1) (only transverse excitations can be measured at this \textbf{Q}-position), but both longitudinal and transverse magnetic signals may be measured at \textbf{Q}=(1,0,0). In our previous treatment, two longitudinal magnetic contributions were detected at \textbf{Q$_0$}=(1,0,0) at low temperature (below T$_0$): a well-defined excitation and a broad quasi-elastic signal interpreted as a continuum of magnetic excitations. The longitudinal character of these two excitations was proved by polarized inelastic scattering performed on IN22 (see measurement in reference~\cite{bourdarot:2010}). No transverse signals, neither elastic nor inelastic have been detected during this experiment. Their temperature dependences were obtained on IN12 with an unpolarized neutron scattering configuration and a final energy E$_f$=4.7~meV.

\begin{figure}[!h]
\begin{minipage}{.475\linewidth}
\begin{center}
\includegraphics[width=66mm]{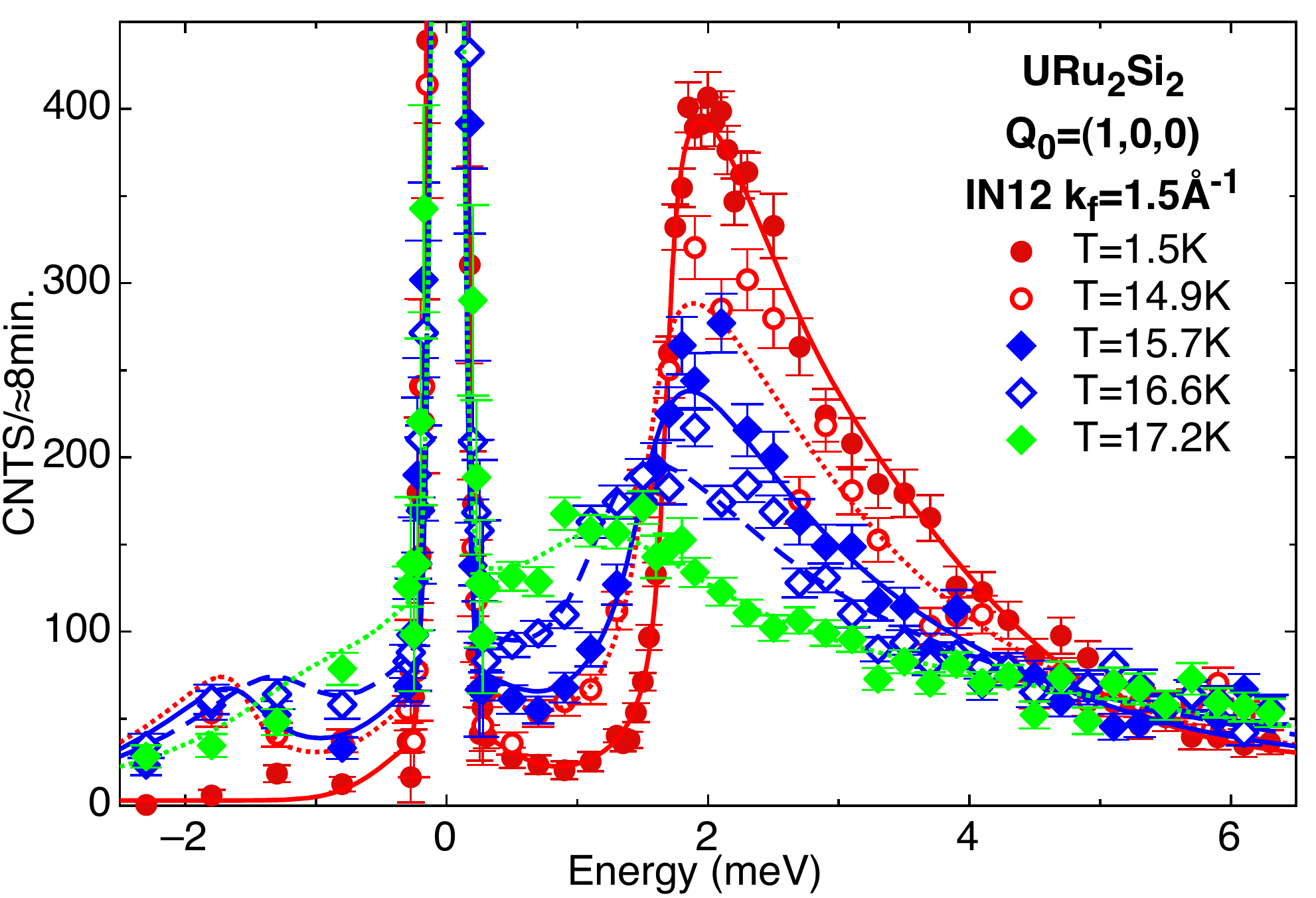}
\caption{Inelastic scattering spectra measured at \textbf{Q$_0$} for temperatures lower than T$_0$.}
\label{Q0BT}
\end{center}
\end{minipage}
\hfill
\begin{minipage}{.475\linewidth}
\begin{center}
\includegraphics[width=66mm]{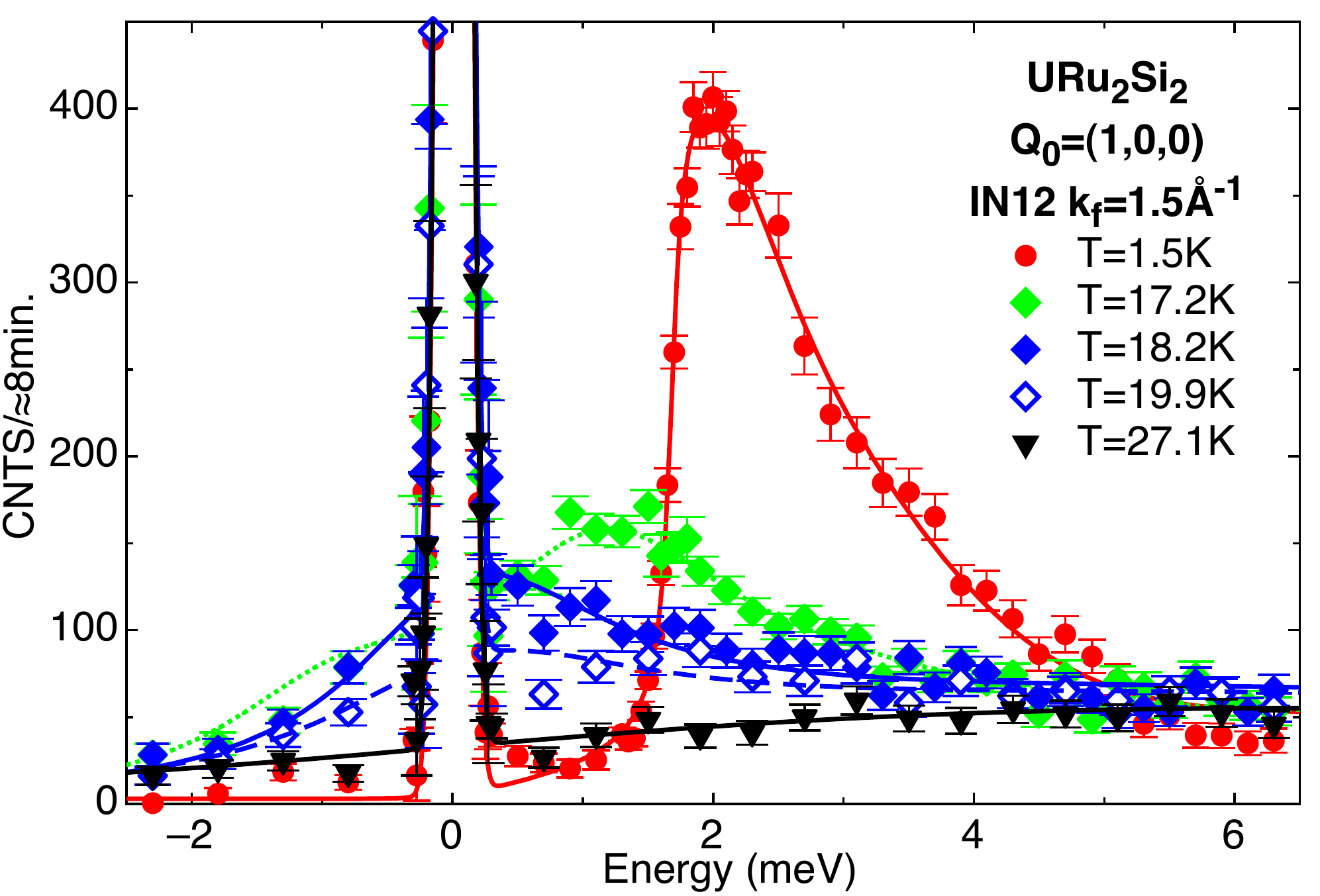}
\caption{Inelastic scattering spectra measured at \textbf{Q$_0$} for temperatures around and above T$_0$.}
\label{Q0HT}
\end{center}
\end{minipage}
\end{figure}

\begin{figure}[!h]
\begin{center}
\includegraphics[width=100mm]{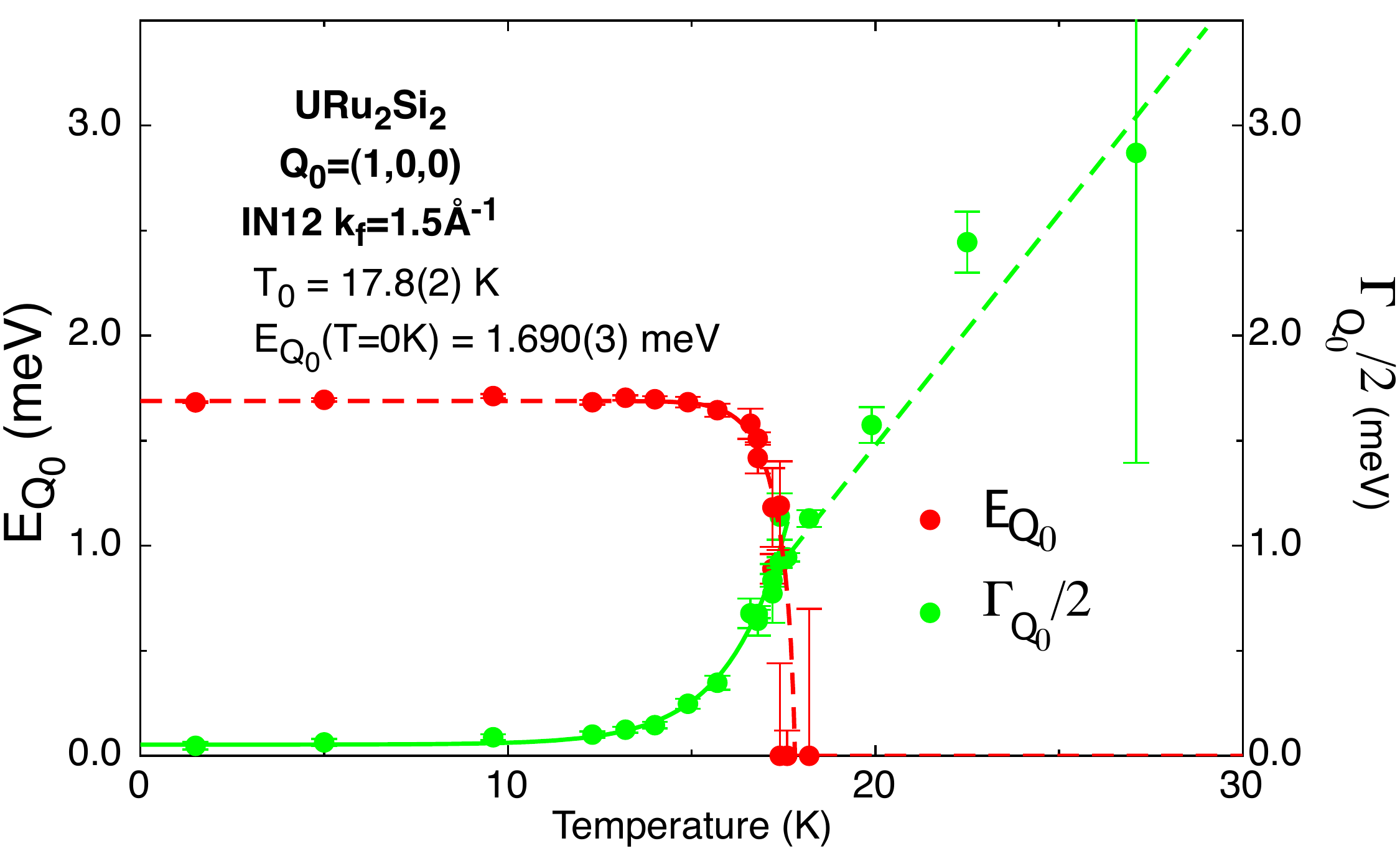}
\caption{Temperature dependence of the energy gap and of the width of the magnetic excitation at \textbf{Q$_0$}.}
\label{Q0gap}
\end{center}
\end{figure}

The inelastic scattering spectra measured below and around T$_0$ are depicted in figures~\ref{Q0BT} and \ref{Q0HT} respectively. Using the damped magnetic excitation function (see appendix~\ref{an3}), the temperature dependences of the energy gap and of its width have been deduced from the inelastic scattering spectra. They are presented in figure~\ref{Q0gap}. These evolutions seem to indicate that the energy gap goes to zero as shown by the spectra close to T$_0$ and only a quite broad quasi-elastic signal persists at high temperature. It is important to note that using the damped magnetic excitation model, it is not necessary to consider another contribution as it was done using the damped harmonic oscillator\footnote[1]{In the Ref.~\cite{bourdarot:2010}, we used the damped harmonic oscillator plus a quasi-elastic functions contribution instead of the damped magnetic excitation function.}. Owing to the fact that the half-width in energy of the excitation becomes larger than the gap for temperatures close to and above T$_0$, there is still the possibility of existence of an energy gap for the excitation above T$_0$. However, the decrease of this gap from 1.7~meV to $\simeq$1.0~meV between 15K to 17K is not an artifact as it can be easily seen on the raw data (figures~\ref{Q0BT} and \ref{Q0HT}).

In conclusion, without any doubt we prove that the energy gap decreases rapidly when approaching T$_0$ from below but it is beyond any treatment to prove whether it persists or not above T$_0$. However, the rapid decrease of the gap seems to indicate the vanishing of this gap at least at T$_0$. Using the damped magnetic excitation model, we also prove that the long-tail magnetic contribution comes directly from the energy dependence of the damped magnetic excitation and has nothing to do with a continuum of magnetic excitations.


\subsection{Q$_1$=(1.4,0,0) magnetic excitation.}

\begin{figure}[!h]
\begin{minipage}{.475\linewidth}
\begin{center}
\includegraphics[width=66mm]{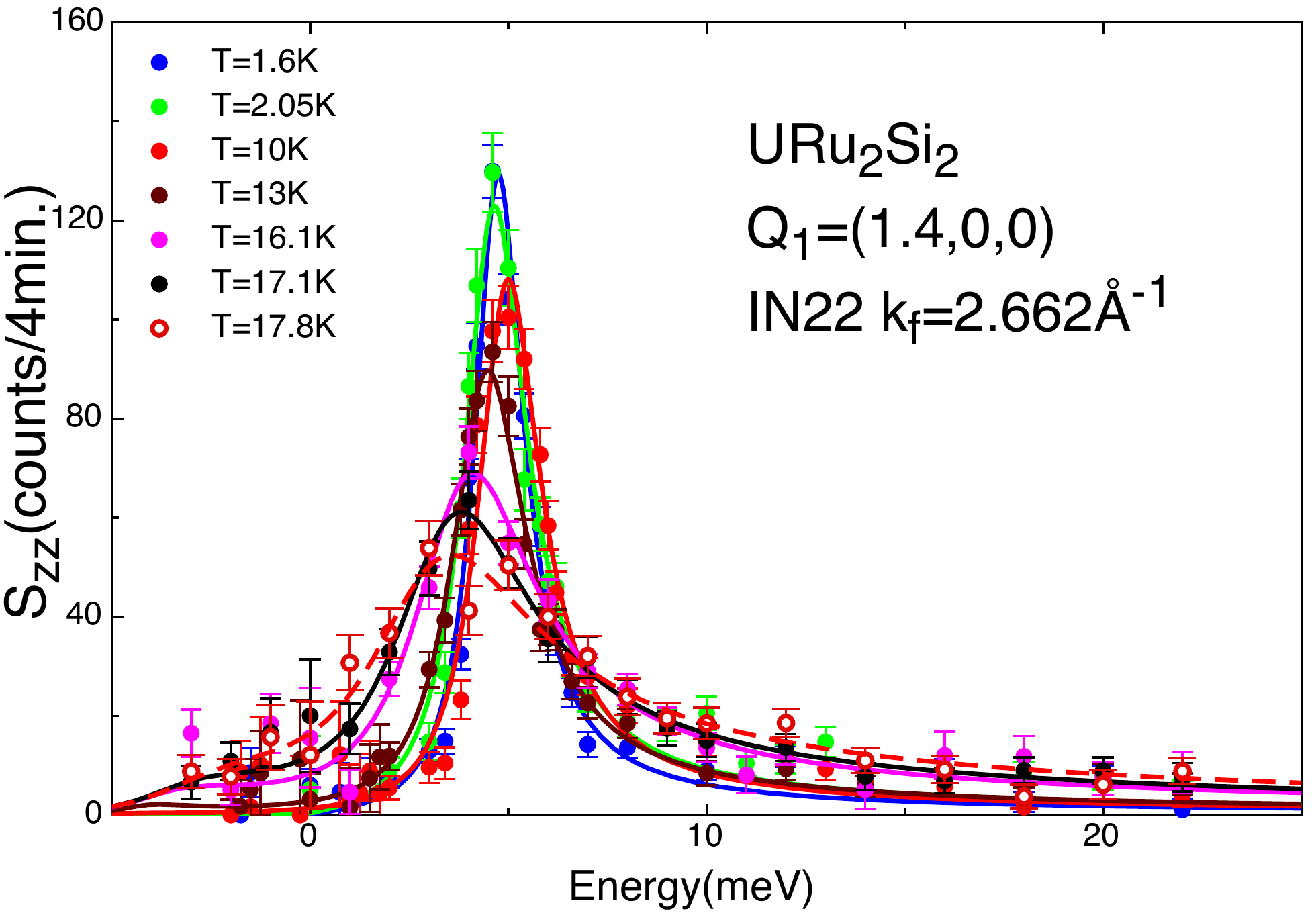}
\caption{Inelastic scattering spectra measured at \textbf{Q$_1$} for temperatures lower than T$_0$. The definition of $S_{zz}$ and the neutrons polarization are given in the note\footnotemark[2].}
\label{Q1BT}
\end{center}
\end{minipage}
\hfill
\begin{minipage}{.475\linewidth}
\begin{center}
\includegraphics[width=66mm]{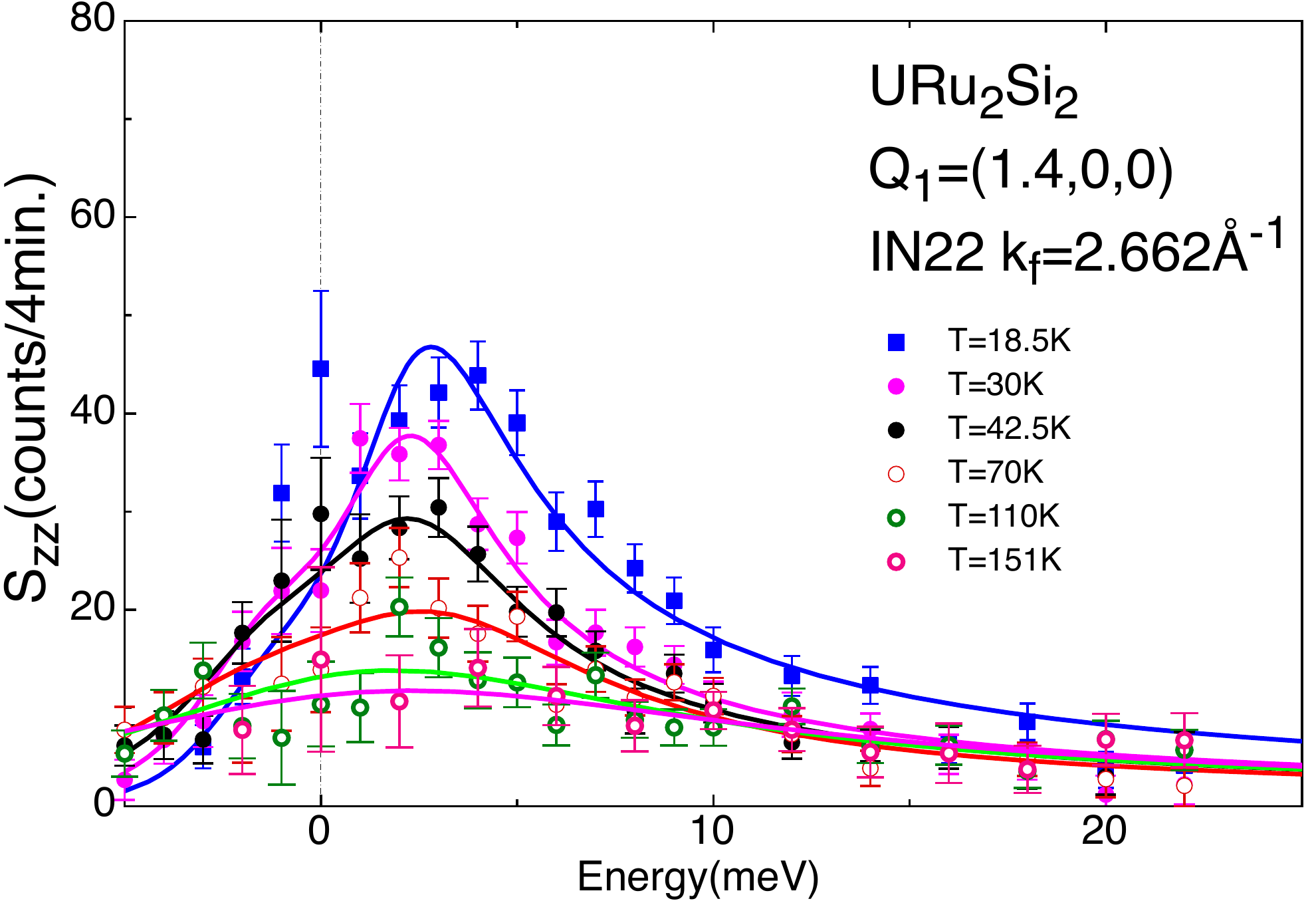}
\caption{Inelastic scattering spectra measured at \textbf{Q$_1$} for temperatures around and above T$_0$. The definition of $S_{zz}$ and the neutrons polarization are given in the note\footnotemark[2].}
\label{Q1HT}
\end{center}
\end{minipage}
\end{figure}

\footnotetext[2]{A high quality single crystal of \urs\ with a rod shape of 8mm high and 4mm diameter with $\vec{b}$ as vertical axis, grown by the Czochralski method in tetra-arc furnace, already used for the inelastic neutron scattering measurements at the wave-vector \textbf{Q$_0$} was installed on the thermal triple-axis IN22 CEA-CRG at ILL. The spectrometer was set-up in its polarized configuration. The experiment was performed with a fixed final energies of 14.7 meV ($k_f$=2.662 \AA$^{-1}$). The beam was polarized by a Heusler monochromator vertically focusing and analyzed in energy, and polarization by a Heusler analyzer vertically and horizontally focusing. The flipping ratio was around 17 and the energy resolution was 0.95 meV. No collimation was installed. The background was reduced by slits placed before and after the sample. The inelastic scans were performed with $\bf{Q}$ parallel to the $\bf{a}$-axis in the non-spin-flip channel which provides all the necessary information (See table~\ref{TabQxy} in annex~\ref{npol}). The intensities of the scans were corrected of monitor error due to harmonic wavelength neutron contamination. 
The difference of scans gives the dynamical magnetic response (not corrected of the Bose factor) with for example $S_{yy}\propto (I^b_{NSF}-I^a_{NSF})$ the transverse susceptibility and $S_{zz}\propto (I^c_{NSF}-I^a_{NSF})$ the longitudinal susceptibility.}

The experimental conditions of this new experiment are described in the footnote\footnotemark[2]. 
As for the magnetic excitation measured at the wave-vector \textbf{Q$_0$}, it was proved that  at low temperature the magnetic excitation at the wave-vector \textbf{Q$_1$} measured by polarized neutron scattering is only longitudinal~\cite{bourdarot:2010}. Magnetic inelastic spectra have been obtained for temperature below (figure~\ref{Q1BT}) and above T$_0$ (figure~\ref{Q1HT}). Contrary to the spectra obtained at the wave-vector \textbf{Q$_0$}, the data shown correspond only to the longitudinal magnetic contributions, thank to the used of polarized neutrons. In practice, these figures display the difference of inelastic scattering spectra obtained with different neutron polarized states (see appendix~\ref{npol}). Only intensities coming from the longitudinal susceptibility ($S_{zz}$ corresponds to this susceptibility without correction of the Bose thermal factor) are detected as it can be seen in the figures~\ref{Q1BT} and~\ref{Q1HT}. From the figure~\ref{Q1HT}, it can be seen as well that the magnetic excitation at \textbf{Q$_1$} has still an open energy gap above T$_0$.

\begin{figure}[!h]
\begin{center}
\includegraphics[width=100mm]{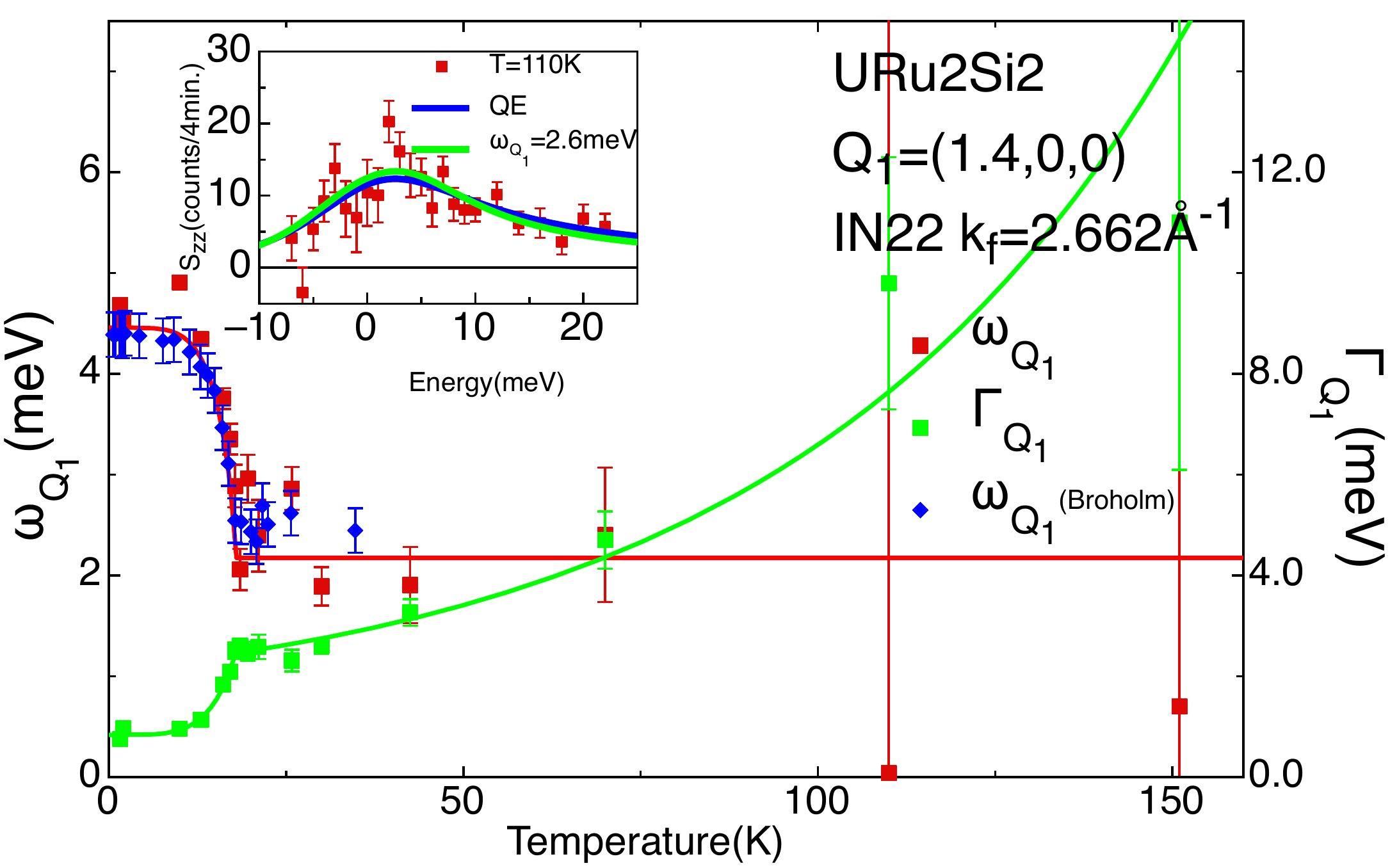}
\caption{Temperature dependence of the energy gap  of the magnetic excitation at \textbf{Q$_1$}. The inset shows the difference of the fits with a quasi-elastic function and a broad inelastic function.}
\label{Q1gapwidth}
\end{center}
\end{figure}

The thermal dependence of the energy gap and of its width in energy at \textbf{Q$_1$} are shown in the figure~\ref{Q1gapwidth}. In agreement with  Broholm measurements~\cite{Broholm:1987,Broholm:1991}, the energy gap does not close at T$_0$ and persists up to at least 70K. In decreasing temperatures close to T$_0$, it increases rapidly from 2.2~meV to 4.2~meV, at low temperature. Above T$\simeq$70K, as for \textbf{Q$_0$}, due to the over-damped character of the excitations, it is impossible to state if the energy gap vanishes or not, as supported in the inset of the figure~\ref{Q1gapwidth} where two fits performed respectively for quasi-elastic and inelastic models  are indistinguishable. This behavior is different from the behavior at \textbf{Q$_0$} for which the thermal dependence of the energy gap indicates a vanishing of the energy gap.

A comparison of temperature dependences of widths in energy indicates that  $\Gamma_\textbf{Q$_0$}$ is larger than $\Gamma_\textbf{Q$_1$}$ (figures~\ref{Q0gap} and~\ref{Q1gapwidth}). The persistence of an energy gap at \textbf{Q$_1$} and the different widths at \textbf{Q$_0$} and \textbf{Q$_1$} seem to indicate a different origin of these two excitations even if the energy gaps have almost an identical relative variation with the temperature.

\begin{figure}[!h]
\begin{center}
\includegraphics[width=100mm]{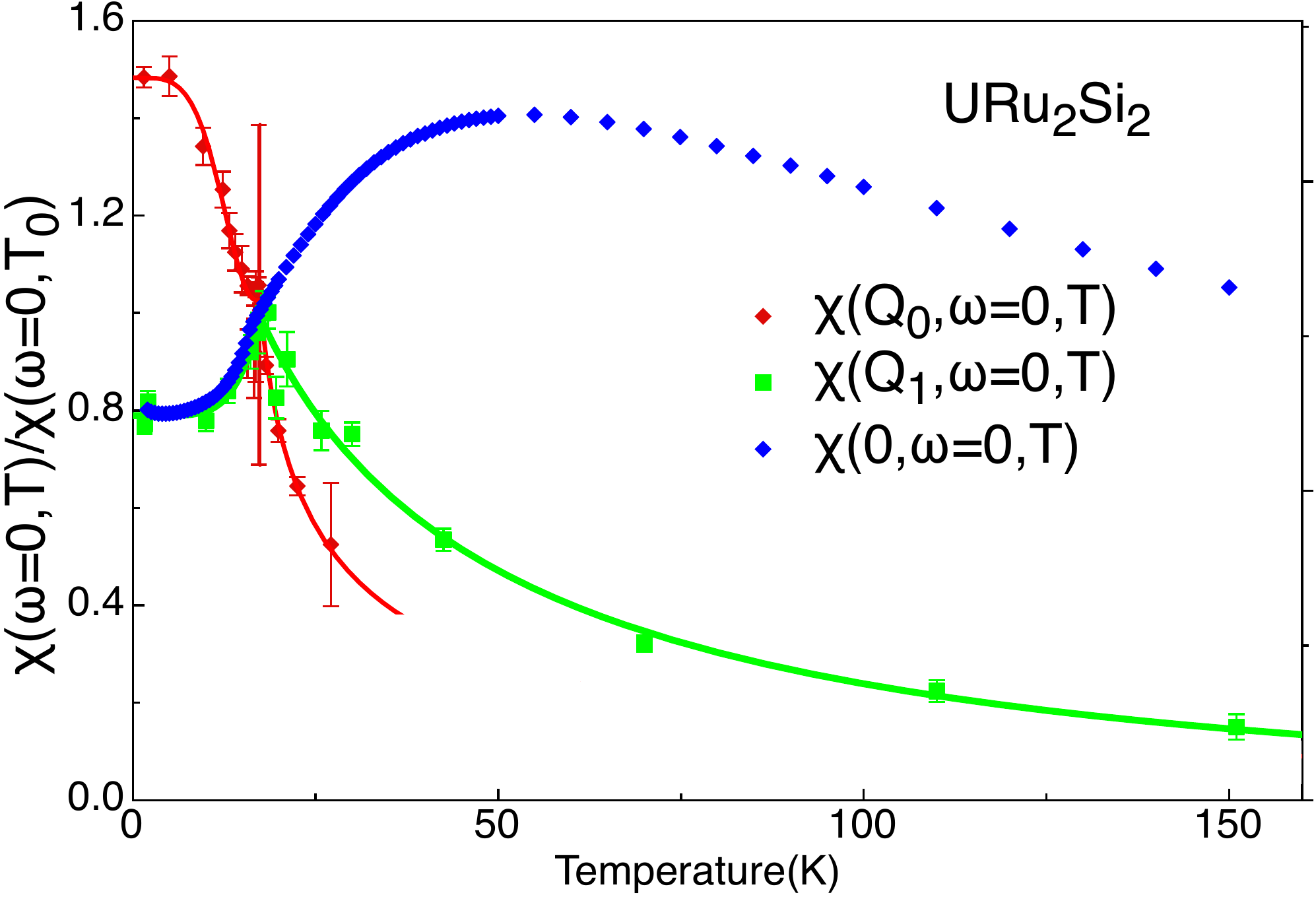}
\caption{Comparison of static susceptibilities $\chi$(Q$_0$,~$\omega$=0,~T) and $\chi$(Q$_1$,~$\omega$=0,~T)  and $\chi$(Q=0,~$\omega$=0,~T), normalized to their values at T$_0$.}
\label{Chi}
\end{center}
\end{figure}

The different origin of these energy gaps is confirmed by the static susceptibilities measured at different  \textbf{Q}-vectors. The thermal variations of the static susceptibilities $\chi(Q,\omega=0,T)$ for \textbf{Q$_0$} and \textbf{Q$_1$} deduced from the analysis with a damped magnetic excitation model show a completely different and unusual behavior as it can be seen on figure~\ref{Chi}. The static susceptibility for \textbf{Q$_0$} decreases as an order parameter, with an inflection point at T$_0$. This variation is unexpected for the static susceptibility at the wave-vector of an antiferromagnetic system, which presents usually in this case a maximum. For \textbf{Q$_1$}, the behavior of the static susceptibility below T$_0$ is similar to the temperature dependence of the static susceptibility measured by conventional magnetometry \hbox{$\chi(\textbf{Q}=0,\omega=0,T)$} (see figure~\ref{Chi}): it presents a small maximum at T$_0$ and decreases at larger temperatures. This high temperature behavior is very different from that of the static susceptibility $\chi(0,\omega=0,T)$ in the same range of temperatures.

\section{Discussion.}

\subsection{The small magnetic moment and the hidden order.}

Although the tiny dipole antiferromagnetic moment m$_0$ can not be the primary order parameter, the moment value of m$_0$ of 0.012(1)$\mu_B$/U is practically sample independent, a fact which might indicate its intrinsic nature. In this case, the tiny moment should better be consider as a secondary order parameter and it may exist a linear coupling between the main order parameter (the so-called hidden order) and m$_0$, which have to belong to the same irreducible representation. Moreover, due to this linear coupling, the properties of a second order magnetic transition are original as demonstrated by N.~Shah~\textit{et al.}~\cite{Shah:2000} and by V.~Mineev\textit{et al.}~\cite{mineev:2005}. Among other predictions, there is the existence of a first order transition between the high pressure antiferromagnetic dipolar state and the hidden order with a transition line ending by a critical point and an unusual magnetic field dependence of the tiny antiferromagnetic moment m$_0$, predicted to present an inflection point for field smaller than the critical field H$_c\simeq$36~T. The field dependence of m$_0$ has been verified experimentally\cite{Bourdarot:2003}, as well as the phase diagram under pressure that shows first order transition from the hidden order state to an antiferromagnetic state, and a ending critical point for the line transition cannot be excluded. The main implication of an intrinsic tiny moment coupled  linearly to the hidden order parameter would be that m$_0$ and the main HO  parameter have to belong to the same irreducible representation $A_2^-$. 

Considering the average value of the four operators belonging to the $A_2^-$ representation with $\{|\Gamma_{t1}^{(1)} \rangle$,$|\Gamma_{t2} \rangle$,$|\Gamma_{t1}^{(2)}\rangle\}$ as lower energy levels, it can be deduces that the hidden order parameter may be the dotriacontapole D$_z^{\beta}$. This multipole differs from the other multipoles of this representation because of a distinctive dependence with the crystal electric field parameter $\theta_1$ ($\theta_1$ is defined in the appendix~\ref{abc}). The dotriacontapole D$_z^{\beta}$ seems also to explain the distribution of magnetization under an applied magnetic field, but the analysis is still under treatment. However, the neutron diffraction experiments cannot prove whether the tiny moment is intrinsic or not. At least, the existence of an intrinsic tiny antiferromagnetic moment m$_0$ seems to contradict some NMR~\cite{Matsuda:2000,Bernal:2001,Bernal:2002, Takagi:2007,Takagi:2012,Kambe:2013a}, NQR~\cite{Saitoh:2005,Bernal:2006} and muon measurements~\cite{Amitsuka:2003}. However, the effects sought by these methods are very small so that any magnetic impurity may give larger signal than the innate response of \urs. Another problem with these measurements is that most of them do not directly probe the magnetic atom but its influence on other atoms or sites. The magnetic influence of a dotriacontapolar moment must be quite peculiar and even if the magnetic field decreases quickly with the distance it should be taken into account for the analysis of these measurements.

Finally, the measurement of the angular dependence of the susceptibility performed by S.~Kambe\cite{Kambe:2013a} seems consistent with the dotriacontapolar moment D$_z^\beta$ coupled with the dipolar moment J$_z$.

\subsection{The magnetic excitations in hidden order phase.}

\begin{figure}[!h]
\begin{center}
\includegraphics[width=100mm]{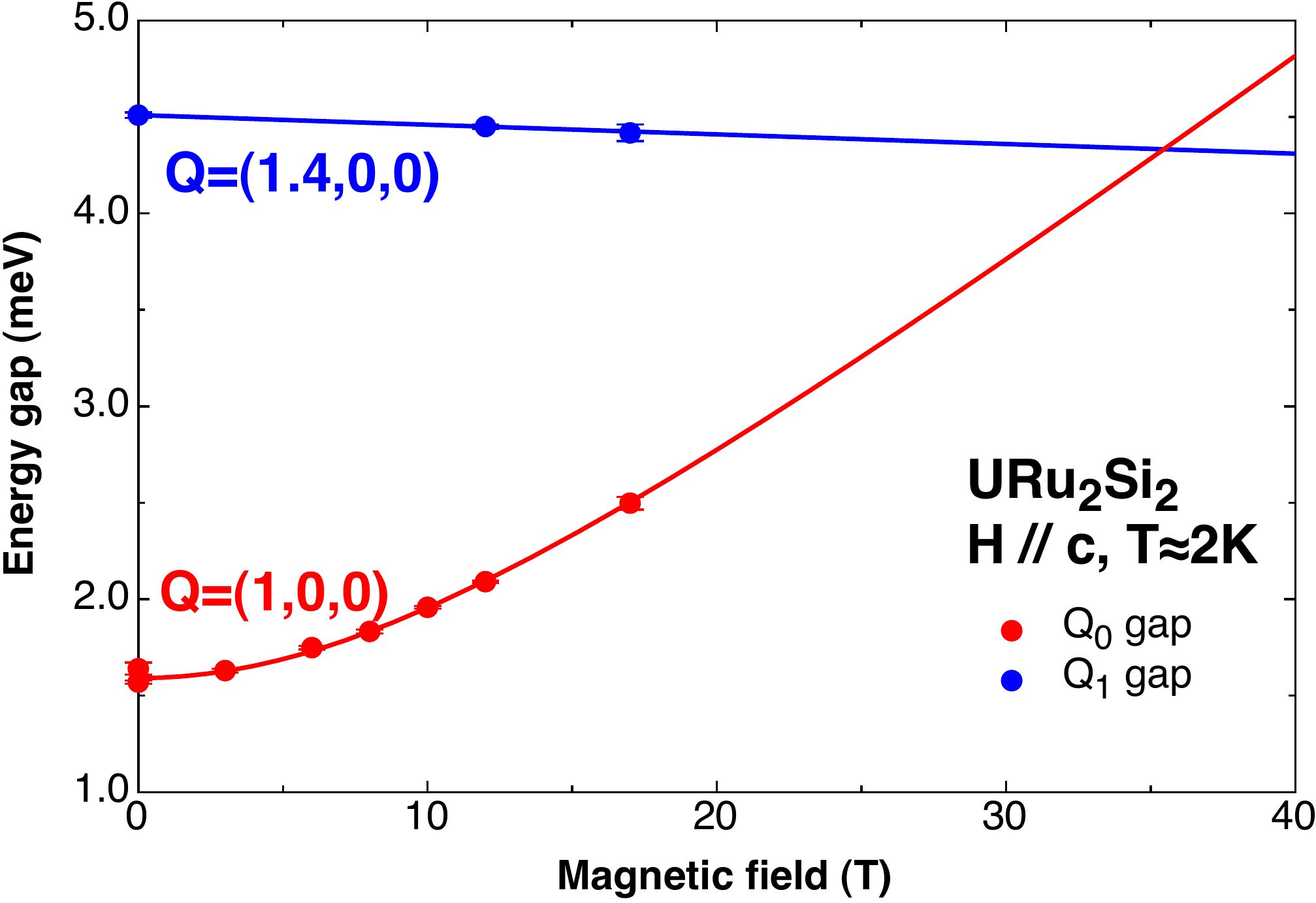}
\caption{Field dependence of the energy gaps at \textbf{Q$_0$} and \textbf{Q$_1$} versus a magnetic field along the $c$-axis.}
\label{GapH}
\end{center}
\end{figure}

\begin{figure}[!h]
\begin{center}
\includegraphics[width=100mm]{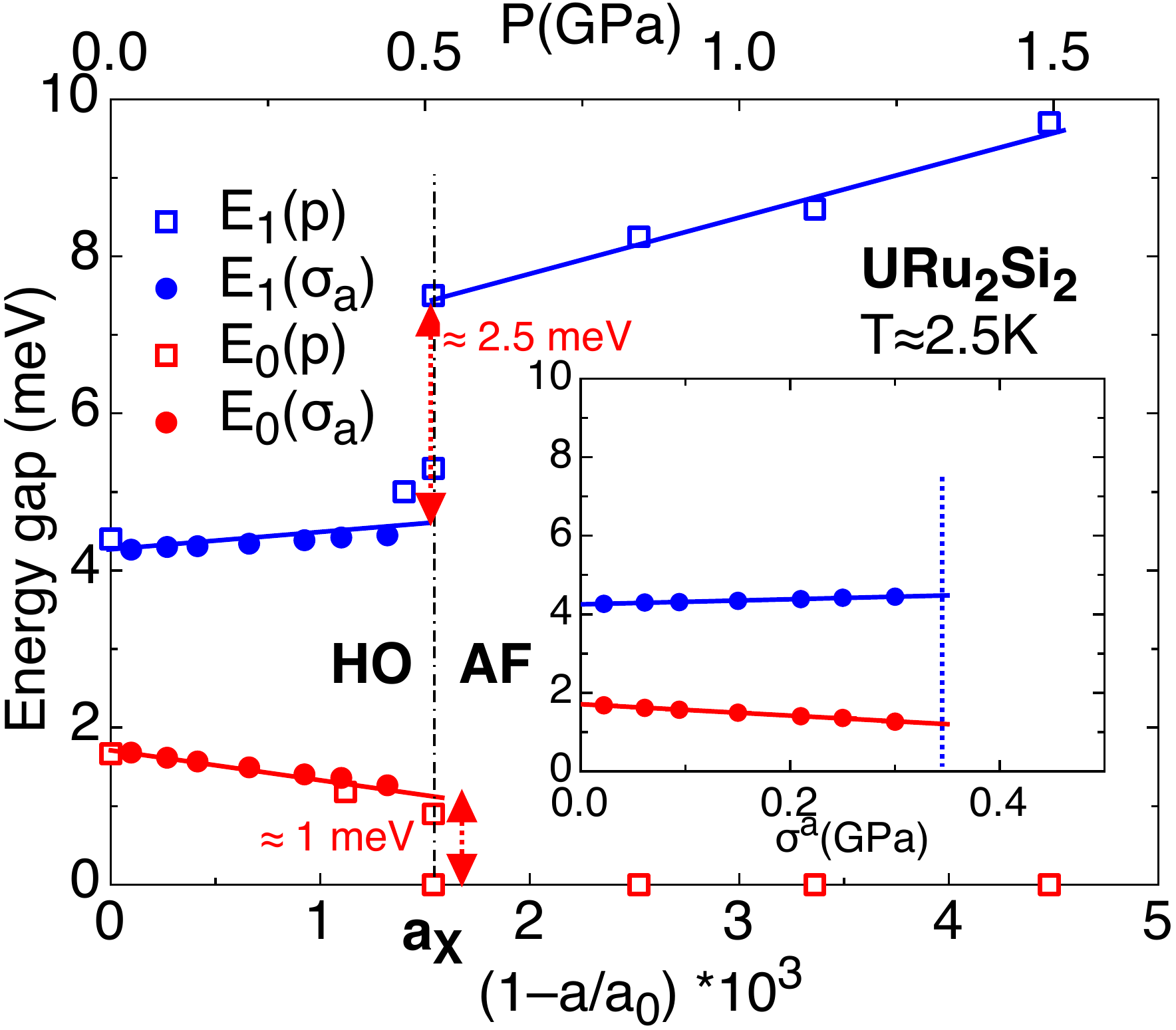}
\caption{Pressure dependence (and uniaxial stress dependence in the inset) of the energy gaps at \textbf{Q$_0$} and \textbf{Q$_1$} versus the variation of the lattice parameter that controls the transition from hidden order state to antiferromagnetic state. The variation of $a$ is obtain either by hydrostatic pressure~\cite{Villaume:2008} or by uniaxial stress~\cite{Bourdarot:2011}.}
\label{GapP}
\end{center}
\end{figure}

\begin{figure}[!h]
\begin{minipage}{.350\linewidth}
\begin{center}
\includegraphics[width=50mm]{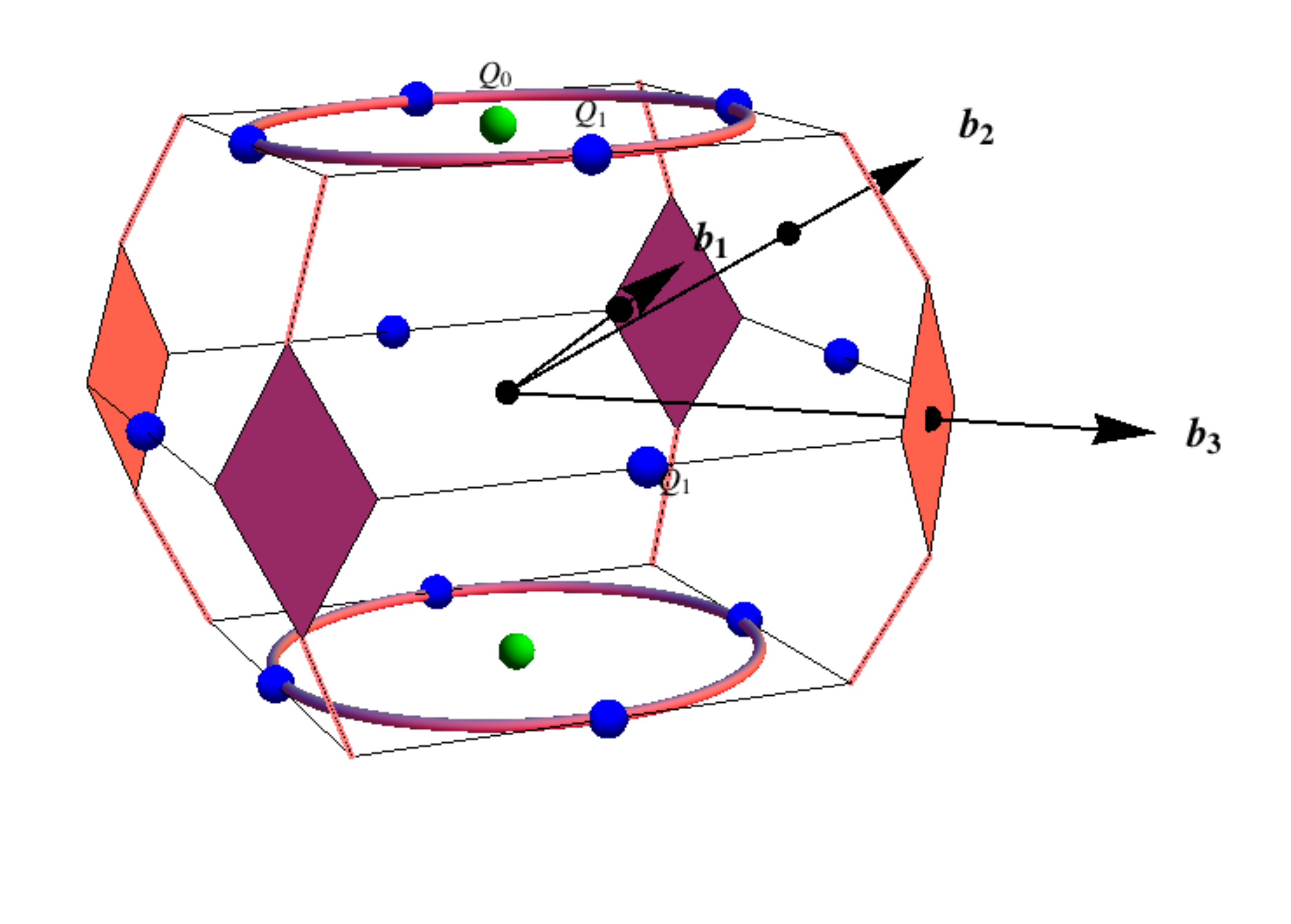}
\caption{Surfaces and parts (colored in orange and purple) of the paramagnetic Brillouin zone with large magnetic excitations at \textbf{Q$_0$}, on a circle centered on \textbf{Q$_0$} and passing through \textbf{Q$_1$}, and on the diamond-shaped surface perpendicular to the vector $\vec{b_3}$=(1/2,1/2,0).}
\label{dismoins}
\end{center}
\end{minipage}
\hfill
\begin{minipage}{.600\linewidth}
\begin{center}
\includegraphics[width=80mm]{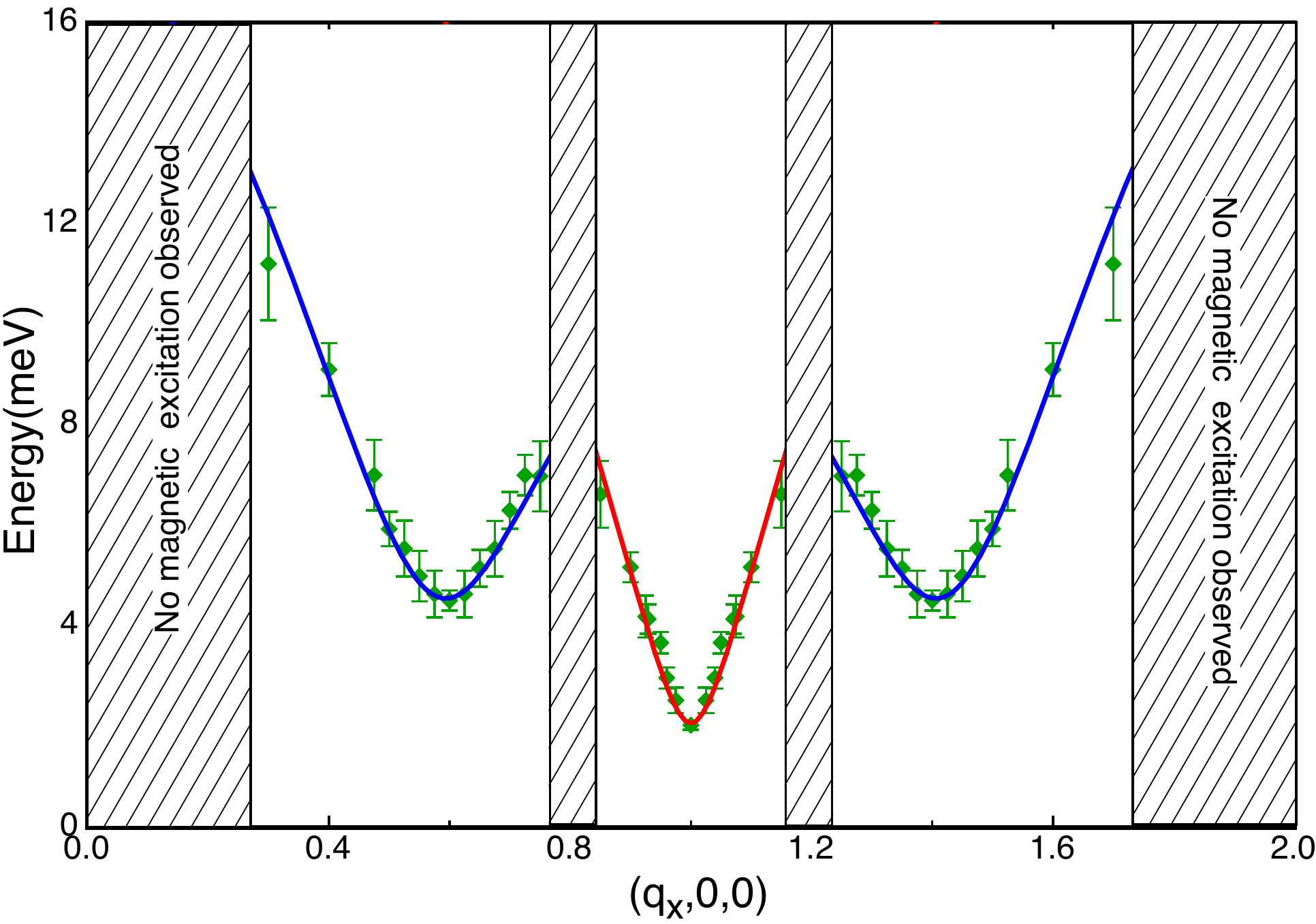}
\caption{New dispersion of the magnetic excitation [h,0,0] at low temperatures taking into account the conclusions reached in this study. Lines are guides for the eyes.}
\label{displus}
\end{center}
\end{minipage}
\end{figure}

The picture according to Broholm \textit{et al.}, which describes the magnetic dispersion at low temperatures from a single mode fitted with seven exchange integrals (see dispersion in references~\cite{Broholm:1987,Broholm:1991} and Wang and Cooper model\cite{Wang:1968,Wang:1969}), seems questionable. Even if the dispersion of magnetic excitations in \urs\ is confirmed at low temperatures around the points \textbf{Q$_0$} and \textbf{Q$_1$}, an important difference appears in our measurements at low temperatures: no signal was detected in the vicinity of the Brillouin zone center ($\Gamma$ point). In fact, the longitudinal magnetic excitations are only present on the surface of the paramagnetic Brillouin zone: at \textbf{Q$_0$}, on a circle centered on \textbf{Q$_0$} and passing through \textbf{Q$_1$}, these two positions corresponding to minimums of the dispersion, and finally on the diamond-shaped  surface perpendicular to the vector $\vec{b_3}$=(1/2,1/2,0) of the Brillouin zone, which presents a maximum around 10~meV in the dispersion (see figure~\ref{dismoins}). The temperature dependence of the excitations at the \textbf{Q$_0$} and \textbf{Q$_1$} points, reveal an inconsistency with a single-excitation model. Above T$_0$, the signal at \textbf{Q$_0$} becomes quasi-elastic while at \textbf{Q$_1$} it remains inelastic. The dynamical  susceptibilities are also very different: slight maximum for \textbf{Q$_1$} at T$_0$, while \textbf{Q$_0$} susceptibility presents an inflection point and continues to increase when the temperatures decrease to 0~K. Differences remain under pressure and magnetic field. The variation with a magnetic field along the $c$-axis of both energy gaps are opposite, with an increase of the energy gap for the excitation at \textbf{Q$_0$} while the energy gap at \textbf{Q$_1$} decreases\cite{Bourdarot:2003} (Figure~\ref{GapH}). The pressure variation of both energy gaps is also opposite but now the energy gap of the excitation at \textbf{Q$_0$} decreases with the pressure before reaching the critical pressure P$_X$ at which \urs\ enters into the antiferromagnetic phase with the disappearance of the excitation at \textbf{Q$_0$}\cite{Villaume:2008}. Conversely, the energy gap of the magnetic excitation at \textbf{Q$_1$} increases with pressure in the HO state and persists in the antiferromagnetic state with a jump at P$_X$ from 5~meV to 8~meV, still increasing in this state, as it is shown in the figure~\ref{GapP}. The persistence of the excitation at \textbf{Q$_1$} in the antiferromagnetic state induced by pressure confirms again that the nature these two excitations is different. However, it is likely that the modification of the magnetic properties above T$_0$ is due to the magnetic excitation at \textbf{Q$_1$} which is essential for the appearance of the magnetic excitation at \textbf{Q$_0$}. Finally, the features of the excitations in the antiferromagnetic phase under pressure indicate that although the magnetic structure becomes primitive tetragonal, the excitation at \textbf{Q$_1$} does not seem to follow the periodicity of this magnetic structure described with the magnetic order wave-vector \textbf{Q}=(0,0,1): no excitations are detected at the position \textbf{Q}=(0.4,0,0) equivalent to \textbf{Q$_1$}. The last remark and the different nature of the excitations at \textbf{Q$_0$} and \textbf{Q$_1$} go in the direction of a dispersion composed of two independent modes, as modeled in figure~\ref{displus}. One mode at \textbf{Q$_0$} with the antiferromagnetic periodicity and another mode at \textbf{Q$_1$} with the paramagnetic periodicity.

Nevertheless, the appearance of a gap in the excitation at \textbf{Q$_0$} below T$_0$ when \urs\ enters in the hidden order state proves that \textbf{Q$_0$} is the wave-vector controlling the hidden order phase as well as the antiferromagnetic phase. On the contrary, the persistence of the excitation at \textbf{Q$_1$} in the hidden order state, antiferromagnetic state under pressure, and in the paramagnetic phase appears as the excitation at \textbf{Q$_1$} is not connected to the appearance of the hidden order parameter. The existence or the appearance at high temperatures of the gapped mode at \textbf{Q$_1$} is certainly essential for the system not to have to end-up in a paramagnetic ground state. This assumption should be also verified in the isomorphic paramagnetic compounds UFe$_2$Si$_2$ and UOs$_2$Si$_2$.

\begin{figure}[!h]
\begin{center}
\includegraphics[width=100mm]{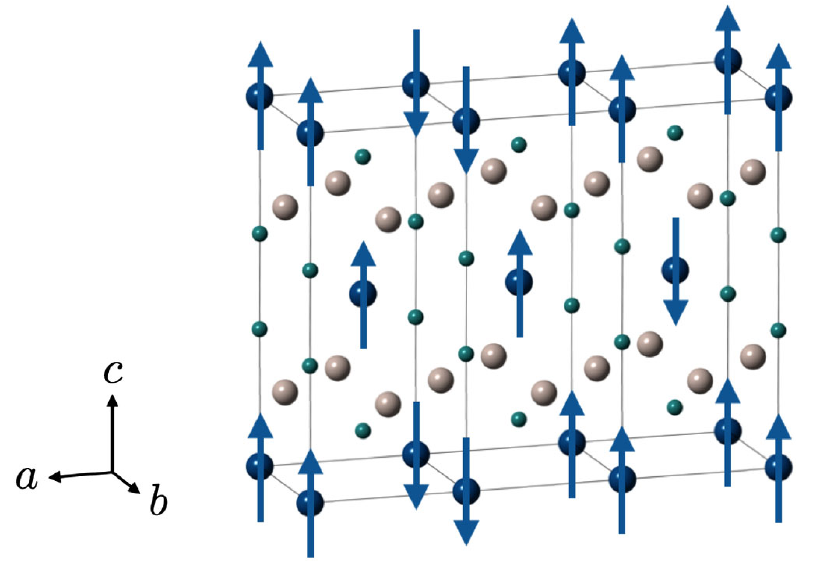}
\caption{Magnetic structure of the field-induced ferrimagnetic phase of \urhh\cite{Kuwahara:2013}.}
\label{highmag}
\end{center}
\end{figure}

Under magnetic field and at low temperatures, the energy gap at \textbf{Q$_1$} tends to become the lowest-energy mode of \urs\ as indicated by the evolution of energy gaps at \textbf{Q$_0$} and at \textbf{Q$_1$} as a function of magnetic field~\cite{Santini:2000,Bourdarot:2003} (see figure~\ref{GapH}). Of course it is impossible to study these excitations by inelastic neutron scattering in magnetic-field values close to H$_c \simeq$36~T. However, recent diffraction results seem to go in the direction of a stabilization of \textbf{Q$_1$} as the wave-vector of the magnetic structure at high magnetic field~\cite{Kuwahara:2013}. In \urhh, for magnetic fields larger than the critical field of H$_c$$\simeq$~26T where this compound enters a ferrimagnetic state, it has been shown from pulsed-field diffraction measurement up to 30T that the magnetic propagation vector of the ferrimagnetic structure, in addition to the ferromagnetic component, is \textbf{k}=(2/3,0,0), close to \textbf{Q$_1$}. This magnetic structure is drawn in the figure~\ref{highmag}.

\section{Conclusion.}

Neutron scattering experiments have not reveal so far the hidden order parameter but they have given some hints. The most striking result is obtained by the measurement of the magnetic form factor that shows modification of some localized polarized electrons between the paramagnetic state and the hidden order state. This result, even if the analysis needs some improvements, indicates that the hidden order seems to be connected to localized electrons. The other point that needs to be clarified is the persistence but also the constant value of the tiny antiferromagnetic moment m$_0$. Is this magnetic phase intrinsic? Neutron scattering cannot give a clear-cut answer and the analysis of results obtained by other techniques may also be not definitive or unambiguous as in their modelings the hidden order is not take into account.

The study of \urs\ by neutron scattering is far from being finished. From the symmetry analysis made in the article~of Khalyavin\cite{Khalyavin:2014}, it may be interesting to study the appearance of domains when a magnetic field is applied along the $c$-axis. Another measurement that may bring some information would be to obtain the magnetization density map under magnetic fields applied along the $a$-axis. But certainly the most promising information will be obtained by the study of the magnetic excitation dispersion in the paramagnetic state and in the  pressure-stabilized antiferromagnetic state. 

The hidden order is undeniably a new type of order which explains why it has not so far been identified. The analysis of the various neutron scattering measurements and those from other techniques are still in progress and their results will likely lead to a correct description of the hidden order phase, even if it may still be sometime before the hidden order is identified.

\section*{Acknowledgement}

We thank D. Aoki, E. Ressouche, R. Ballou, J. Flouquet, J.~P. Sanchez and J.~P.~Brison for helpful discussions.

\section*{Funding}
Part of this work was supported by Euromagnet II via the EU under Contract No. RII3-CT-2004-506239, ERC starting grant (NewHeavyFermion), and French ANR projects (CORMAT, SINUS, MAGFINS).

\appendix

\section{Crystalline electric field in \urs.}
\label{Ann2}

Crystal field levels of the uranium ion U$^{4+}$ in the fundamental multiplet $^3H_4$ in a tetragonal symmetry.

\begin{table}[b]
  \centering \begin{tabular}{|c|c|c|c|}
\hline
 Level  & m & $|\Psi\rangle$ & Energy\\
\hline
$|\Gamma^{(1)}_{t1}\rangle$ & 1 & $ \frac{\sin(\theta_1)}{\sqrt{2}}(|-4\rangle+|+4\rangle)+\cos(\theta_1)|0\rangle$ & $d -\sqrt{f^2+k^2}$ \\
$|\Gamma_{t2}\rangle$          & 1 & $ \frac{1}{\sqrt{2}}(|+4\rangle-|-4\rangle)$ & $d+f$ \\
$|\Gamma^{(2)}_{t1}\rangle$ & 1 & $ \frac{\cos(\theta_1)}{\sqrt{2}}(|-4\rangle+|+4\rangle) -\sin(\theta_1)|0\rangle$ & $d + \sqrt{f^2+k^2}$\\
\hline
$|\Gamma_{t3}\rangle$          & 1 & $ \frac{1}{\sqrt{2}}(|-2\rangle+|+2\rangle)$ & $b - |g|$\\
$|\Gamma_{t4}\rangle$       & 1 & $ \frac{1}{\sqrt{2}}(|+2\rangle - |-2\rangle)$ & $b + |g|$\\
\hline
$|\Gamma^{(1)}_{t5}\rangle$ & 2 & $\sin(\theta_2)|\pm 3\rangle+\cos(\theta_2)|\mp 1\rangle$ & $c +\sqrt{e^2+h^2}$\\
$|\Gamma^{(2)}_{t5}\rangle$ & 2 & $ \cos(\theta_2)|\pm 3\rangle - \sin(\theta_2)|\mp 1\rangle$ & $c -\sqrt{e^2+h^2}$\\
\hline
\end{tabular}
  \caption{Crystal field levels of the fundamental multiplet $^3H_4$ of U$^{4+}$ in a tetragonal structure. The coefficients $a$, $b$, $c$, $d$, $e$, $f$, $g$, $h$, et $k$ as well as $\cos(\theta_i)$ and $\sin(\theta_i)$ are defined in the table~\ref{abc}.}
  \label{Tab3H4}
\end{table}

\begin{table}[b]
  \centering 
\begin{tabular}{cc|cc}
$a$  = &  $28 B_2^0 + 840 B_4^0  + 5040 B_6^0$ &\\
$b $ = &  $-8 B_2^0 -660 B_4^0  + 27720 B_6^0$  & $\cos(\theta_1)=\frac{\gamma}{\sqrt{1+\gamma^2}}$  &  $\sin(\theta_1)=\frac{1}{\sqrt{1+\gamma^2}}$ \\ 
$c  $= & $-5 B_2^0 -360 B_4^0  - 10080 B_6^0 $ \\
$d  $= & $4 B_2^0 +960 B_4^0  - 10080 B_6^0 $ & $\gamma = \frac{-f +\sqrt{f^2+k^2}}{k}$ & $  \pi > \theta_1 \geqslant 0$\\
$e  $= &  $12 B_2^0 -900 B_4^0  -11340 B_6^0$  \\
$f  $= &  $24B_2^0 -120 B_4^0  +15120 B_6^0 $ & $\cos(\theta_2)=\frac{\alpha}{\sqrt{\alpha^2 + 1}}$  &  $\sin(\theta_2)=\frac{1}{\sqrt{\alpha^2 + 1}}$ \\
$g $ = & $-180 B_4^4 +2520 B_6^4$ & \\
$h  $= &$ ( 60 B_4^4 -180  B_6^4)\sqrt{7}$ &$\alpha= \frac{-e+\sqrt{e^2+h^2}}{h}$ & $ \pi > \theta_2 \geqslant 0$\\
$k  $= &$ ( 24 B_4^4  + 720 B_6^4)\sqrt{35}$ & \\
\end{tabular}
\caption{Coefficients of the crystal electric field versus the Steven's parameters $B_l^m$ and the definitions of $\theta_1$ and $\theta_2$.}
\label{abc}
\end{table}

\begin{table}[h]
\centering
\begin{tabular}{||c||c|c|c||}
\hhline{|t:=:t:===:t|}
Rk 1$-$5 & $|\Gamma^{(1)}_{t1}\rangle$ & $|\Gamma_{t2}\rangle$ & $|\Gamma^{(2)}_{t1}\rangle$ \\
\hhline{|b:=:b:===:b|}
\hhline{|t:=:t:===:t|}
$|\Gamma^{(1)}_{t1}\rangle$ & $O_2^0$,\:$H_0$,\:$H_4$ & $J_z$,\:$T^\alpha_z$,\:$\imath H_z^\alpha$,\:$D_z^\alpha$,\:$D_z^\beta$ & $O_2^0$,\:$H_0$,\:$H_4$,\:$-\imath D_{s4}$ \\
\hhline{||-||---||}
$|\Gamma_{t2}\rangle$& $J_z$,\:$T^\alpha_z$,\:$-\imath H_z^\alpha$,\:$D_z^\alpha$,\:$D_z^\beta$ & $O_2^0$,\:$H_0$ & $J_z$,\:$T^\alpha_z$,\:$\imath H_z^\alpha$,\:$D_z^\alpha$,\:$D_z^\beta$ \\
\hhline{||-||---||}
$|\Gamma^{(2)}_{t1}\rangle$ & $O_2^0$,\:$H_0$,\:$H_4$,\:$\imath D_{s4}$ &  $J_z$,\:$T^\alpha_z$,\:$-\imath H_z^\alpha$,\:$D_z^\alpha$,\:$D_z^\beta$ & $O_2^0$,\:$H_0$,\:$H_4$ \\
\hhline{|b:=:b:===:b|}
\end{tabular}
\caption{Nonzero Matrix elements for magnetic multipoles up to the fifth rank for J=4 in a tetragonal symmetry. Transition elements $T_{xyz}$, $T^\alpha_x$, $T^\alpha_y$, $T^\beta_x$, $T^\beta_y$, $T^\beta_z$, $H_2$, $H_x^\alpha$, $H_y^\alpha$, $H_x^\beta$, $H_y^\beta$, $H_z^\beta$, $D_{s2}$, $D_x^\alpha$, $D_y^\alpha$ $D_x^\beta$, $D_y^\beta$, $D_x^\gamma$, $D_y^\gamma$, and $D_z^\gamma$ do not appear as transition elements for the three singlets $|\Gamma^{(1)}_{t1}\rangle$, $|\Gamma^{(2)}_{t1}\rangle$ and $|\Gamma_{t2}\rangle$.}
\label{Rk3}
\label{zxc}
\end{table}

\section{Dynamical susceptibilities.}
\label{an3}
\subsection{Damped harmonic oscillator.}

For phonons, it is well known that the excitation is fitted by a damped harmonic oscillator with an imaginary part of the dynamical susceptibility is given then by: \\
-If $\gamma/2 < \omega_0$, 
\begin{equation}
\label{AnIIIeq3}
\begin{split}
\chi''_{xx}(\omega)  = \frac{1}{2m\omega_1} &\left[\frac{\gamma/2}{(\omega-\omega_1)^2+(\gamma/2)^2}\right.\\
&\left.-\frac{\gamma/2}{(\omega+\omega_1)^2+(\gamma/2)^2}\right] 
\end{split}
\end{equation}
\noindent
with $\omega_1=\sqrt{\omega_0^2-(\gamma/2)^2}$, the renormalized resonance and $\gamma/2$, the damping.

-If $\gamma/2 > \omega_0$, the susceptibility corresponds to an exponential damping with a characteristic time $\tau=1/\gamma$.

\begin{equation}
\label{AnIIIeq4}
\chi''(\omega) = \frac{1}{m \gamma}\frac{\omega}{\omega^2+(\omega_0^2/\gamma)^2}
\end{equation}

\subsection{Magnon and magnetic excitations.}
\label{AnIV}

For magnetic excitations, the dynamical susceptibility is much more complicated to obtain and a future publication will be dedicated to this calculation. The resulting fit function is valid for magnon as well as for magnetic excitations in crystal electric field scheme:

\begin{equation}
\label{form66}
\chi_{zz}''(\omega) = \frac{\chi_0}{2} \gamma \omega \left(\frac{1}{(\omega - \omega_1)^2 + \gamma^2} + \frac{1}{(\omega + \omega_1)^2 + \gamma^2} \right)
\end{equation}

\noindent where $\gamma$ is the damping factor. When the damping is small ($\gamma \rightarrow 0$), the expression corresponds to the difference of two Dirac's functions\footnotemark[3]:
$$\lim_{\gamma \rightarrow 0} \chi''_{zz}(\omega) = \pi \omega_1 \frac{\chi_0}{2} \left[\delta(\omega-\omega_1)-\delta(\omega+\omega_1)\right]$$

\footnotetext[3]{The Dirac's function: $\delta(x)=\frac{1}{\pi}\mathrm{Im}\left(\lim\limits_{\eta \rightarrow 0^+}(x-\imath \eta)^{-1}\right)$}

For this model, there is a characteristic time $\tau_c$ and for time smaller than this characteristic time, this approximation is no more valid. This means that there is a cutoff for frequencies larger than $\omega_c=2\pi/\tau_c$ that allows to obtain a convergence of the susceptibility.

The magnetic excitations spectra have to be adjusted with this function that converges naturally to the quasi-elastic function when the gap is going to zero energy:
$$\lim_{ \omega_1 \rightarrow 0} \chi''_{zz}(\omega) = \chi_0 \frac{\omega\,\gamma}{\omega^2 +\gamma^2}$$

\section{Polarized neutron scattering analysis.}
\label{npol}

The magnetic intensities in the spin-flip (SF)/non-spin-flip (NSF) channels are given in the table below. The spectrometer basis is giving by ($x$, $y$, $z$) with $x$ along the scattering vector and $z$ along the vertical axis.  (P$_x$, P$_y$, P$_z$) is the polarization vector in this basis. For the measurement on IN22 with $\vec{Q}\ //\ \vec{a^*}$ and $\vec{c^*}$ vertical, the spin-flip and no-spin-flip intensities are:\\

\begin{table}[h]
\centering
\begin{tabular}{|c|c|c|c|}
\hline
Pol. & Int. &NSF & SF \\
\hline
P$_x$ & I$^a$ &N+bg & S$_{yy}$+S$_{zz}$+bg' \\
\hline
P$_y$ & I$^b$ &N+bg+S$_{yy}$ & S$_{zz}$+bg' \\
\hline
P$_z$ & I$^c$ &N+bg+S$_{zz}$ & S$_{yy}$+bg' \\
\hline
\end{tabular}
\caption{Magnetic intensities contribution with $\vec{Q}$=(Q$_x$,0,0) and $\vec{\mathrm{S}}_{spec.}$=(S$_{xx}$,S$_{yy}$,S$_{zz}$) in the spectrometer basis and the sample with c along the vertical axis.}
\label{TabQxy}
\end{table}

\end{document}